\documentclass[preprint,epsfig,graphicx,showpacs,eqsecnum,aps,amsmath,mathrsfs,eufrak,amssymb,floatfix]{revtex4}
\usepackage{graphicx,dcolumn,rotating,booktabs}
\usepackage{epsfig}
\begin{document}
\title{Analytical description of the Coherent State Model for near vibrational and well deformed nuclei.}

\author{A. A. Raduta$^{a),b)}$, R. Budaca$^{a)}$ and Amand Faessler $^{c)}$}

\address{$^{a)}$ Department of Theoretical Physics, Institute of Physics and
  Nuclear Engineering,POBox MG6, Bucharest 077125, Romania}

\address{$^{b)}$Academy of Romanian Scientists, 54 Splaiul Independentei, Bucharest 050094, Romania}

\address{$^{c)}$Instit\"{u}t f\"{u}r Theoretishe Physik der Universit\"{a}t T\"{u}bingen, Auf der Morgenstelle 14, D-72076 T\"{u}bingen,  Germany}
\renewcommand{\theequation}{2.\arabic{equation}}
\begin{abstract}
Analytical formulas for the excitation energies as well as for the  electric quadrupole reduced transition probabilities in the ground, beta and gamma bands were derived within the coherent state model for the near vibrational and well deformed nuclei. Numerical calculations were performed for 42 nuclei exhibiting various symmetries and therefore with specific properties. Comparison of the calculation results with the corresponding experimental data shows a good agreement. The parameters involved in the proposed model satisfy evident regularities being interpolated by smooth curves. Few of them, which fall out of the curves, are interpreted as signatures for a critical point in a specific phase transition. This is actually supported also by the figures showing the excitation energy dependence on the angular momentum. The formulas provided for energies and B(E2) values are very simple, being written in a compact form, and therefore easy to be handled to explain the new experimental data. 
\end{abstract}
\maketitle

\section{Introduction}
\label{sec:level1}
Since the liquid drop model was developed \cite{Bohr}, the
quadrupole shape coordinates  were widely used  by both
phenomenological and microscopic formalisms to describe the basic properties of
nuclear systems. Based on these coordinates, one defines quadrupole
boson operators in terms of which model Hamiltonians and transition operators
are defined. Since the original spherical harmonic liquid drop model was able
to describe only a small amount of data for spherical nuclei, several
improvements have been added. Thus, the Bohr-Mottelson model was generalized by
Faessler and Greiner\cite{GrFa}
in order to describe the small oscillations around a deformed shape which
results in obtaining a flexible model, called Vibration Rotation Model (VRM),
suitable for the description of deformed nuclei. Later on \cite{Gneus} this picture was
extended by including anharmonicities as low order invariant  polynomials in the
quadrupole coordinates. With a suitable choice of the parameters involved in the model
Hamiltonian the equipotential
energy surface may exhibit several types of minima \cite{Hess} like spherical,
deformed prolate, deformed oblate, deformed triaxial, etc.
 To each equilibrium shape, specific properties for excitation energies and electromagnetic transition
 probabilities show up. Due to this reason,
one customarily says that static values of intrinsic coordinates determine a phase for the
nuclear system. A weak point of the boson description with a complex anharmonic Hamiltonian consists of the large number of the structure parameters which are
to be fitted.  A much smaller number of parameters is used by the
coherent state model (CSM) \cite{Raduta1} which uses a restricted collective space generated through angular momentum projection by three deformed orthogonal functions of coherent type. The model is able to describe in a realistic fashion transitional and well deformed nuclei of various shapes including states of high and very high angular momentum. Various extensions to include other degrees of freedom like isospin
\cite{Rad2}, single particle\cite{Rad3} or octupole degrees\cite{Rad4} of freedom have been formulated\cite{Rad5}.
  
It has been noticed that a given nuclear phase may be
associated to a certain symmetry. Hence, its properties may be described with
the help of the irreducible representation of the respective symmetry group.
Thus, the gamma unstable nuclei can be described by the $O(6)$ symmetry
\cite{Jean}, the gamma triaxial nuclei by the rigid triaxial rotor $D2$ symmetry
\cite{Filip},
the symmetric rotor by the $SU(3)$ symmetry and the spherical vibrator by the
$U(5)$ symmetry.
Thus, even in the 50's, the   symmetry properties have been greatly appreciated.
 However, a big push forward was brought by the interacting boson
 approximation
(IBA) \cite{Iache,Iache1}, which succeeded to describe the basic properties of a large number of
nuclei in terms of the symmetries associated to the system of quadrupole (d) and
monopole (s) bosons  which  generate a $U(6)$ algebra. The three
limiting symmetries $U(5)$, $O(6)$, $SU(3)$ mentioned above, are dynamic symmetries
for $U(6)$. Moreover, for each of these symmetries a specific group reduction
chain  provides the quantum numbers characterizing the states, which are suitable
  for a certain region of nuclei. Besides  the virtue of unifying the group
  theoretical descriptions of nuclei exhibiting different symmetries,
  the procedure defines very simple reference pictures for the limiting cases. For nuclei
 lying close to the region characterized by a certain symmetry,
 the perturbative corrections are to be included. In Refs.\cite{Iache2,Iache9} it was shown that the critical points of some transitions correspond themselves to certain symmetries which may be  described by the solutions of specific differential equations.

Many publications developing the mentioned formalisms as well advancing new approaches have been accumulated along the time. To mention them would take too much space and moreover one meet the risk of omitting involuntary some valuable contribution. Due to these reason we shall mention only those papers which are related to the present work.

The present paper is devoted to a systematic study of the CSM approach. In performing the present investigation, we have been stimulated by our previous publication \cite{RaBuFa} where the ground band energies  for states of angular momentum going up to high values (36) and  for  a large number of nuclei (44), have been described with very high accuracy with very simple and compact formulas. Thus, in a way this work is a natural extension of the procedure of the quoted paper to the excited bands. The exact and complex formulas for the matrix elements of the model Hamiltonian as well as of the E2 transition operator have been expanded in power series of a deformation variable x ( $=d^2$ where $d$ is a real parameter which simulates the nuclear deformation) for small deformation and in power series of $1/x$ when the nuclear deformation is large.
As a result, analytical compact formulas are obtained for both excitation energies and quadrupole electric transition probabilities. These formulas are positively tested for a large number of nuclei.

The above mentioned project was achieved according to the following plan.
The basic ideas of the CSM approach are shortly reviewed in Section II. The near vibrational regime is described  in Section III, while the asymptotic expansion for large deformations is given in Section IV. Numerical applications are presented in Section V and the final conclusions are summarized in Section VI.
\renewcommand{\theequation}{2.\arabic{equation}}
\section{ The coherent state model for three interacting bands}
\label{sec:level2}
The model proposed in Ref.\cite{Raduta1} is known under the name of CSM (coherent state model) and aims at describing in a realistic fashion
the lowest three rotational bands, ground, beta and gamma.
Here we describe briefly the basic ingredients. First one builds up a collective boson space being guided by the experimental picture. Thus, each band is generated by projecting the angular momentum from three orthogonal deformed states which modeled the ground, beta and gamma bands respectively, in the intrinsic frame of reference. The deformed ground band state is an axially deformed coherent state $\psi_g$, while the other two intrinsic states are orthogonal polynomial excitations of the ground band model function. These excitations were chosen such that they are mutually orthogonal both before and after angular momentum projection. All three states are depending on a real parameter $d$ which simulates the nuclear deformation. In the limit $d \to 0$ the projected states must go to the first three highest seniority states of the boson multiplets while in the asymptotic region, i.e. large value for $d$, the states written in the intrinsic frame of reference have expressions similar to those associated to the liquid drop model, in the strong coupling regime. By this requirement we assure that the model states have a behavior which is consistent with the so called Sheline-Sakai scheme \cite{She,Saka} which makes a continuous link between the vibrational and rotational spectra. These properties are satisfied by the following three sets of projected states:
\begin{eqnarray}
\phi_{JM}^{g}(d)&=&N_{J}^{g}P_{M0}^{J}\psi_{g},\,\,\,\,\psi_{g}=\exp{\left[d(b_{0}^{\dagger}-b_{0})\right]}|0\rangle,\\
\phi_{JM}^{\beta}(d)&=&N_{J}^{\beta}P_{M0}^{J}\Omega^{\dagger}_{\beta}\psi_{g},\,\,\,\,\Omega_{\beta}^{\dagger}=\left(b^{\dagger}b^{\dagger}b^{\dagger}\right)_{0}+\frac{3d}{\sqrt{14}}\left(b^{\dagger}b^{\dagger}\right)_{0}-\frac{d^{3}}{\sqrt{70}},\\
\phi_{JM}^{\gamma}(d)&=&N_{J}^{\gamma}P_{M2}^{J}\Omega^{\dagger}_{\gamma,2}\psi_{g},\,\,\,\,\Omega_{\gamma,m}^{\dagger}=\left(b^{\dagger}b^{\dagger}\right)_{2,m}+d\sqrt{\frac{2}{7}}b^{\dagger}_{m}.
\end{eqnarray}
Within the restricted space just defined, one constructs an effective Hamiltonian by requiring a maximal decoupling, i.e. the off diagonal matrix elements are equal or close to zero. Ideally would be to have a diagonal Hamiltonian but this is not possible due to the gamma band. However, one solution is given by the 
 six order quadrupole boson Hamiltonian:
\begin{equation}
H^{(2)}=A_{1}(22\hat{N}+5\Omega_{\beta'}^{\dagger}\Omega_{\beta'})+A_{2}\hat{J}+A_{3}\Omega_{\beta}^{\dagger}\Omega_{\beta},\,\,\,\,\Omega^{\dagger}_{\beta'}=\left(b^{\dagger}b^{\dagger}\right)_{0}+\frac{d^{2}}{\sqrt{5}}.
\end{equation}
where $\hat{N}$ denotes the quadrupole boson number operator:
\begin{equation}
\hat{N}=\sum_{\mu =-2}^{2}.
\end{equation}
Indeed $H^{(2)}$  has the property that its matrix elements between a beta band state and a ground band or a gamma band state are vanishing.

The interaction of the $\beta$-band  with the rest of the boson space might be  simulated by some additional terms in the model Hamiltonian, which do not affect the decoupling feature of the band:
\begin{equation}
\Delta H=A_{4}(\Omega_{\beta}^{\dagger}\Omega_{\beta'}^{2}+h.c.)+A_{5}\Omega_{\beta'}^{\dagger2}\Omega_{\beta'}^{2}.
\end{equation}
Thus, the total Hamiltonian used by CSM for describing the ground, beta and gamma bands has the form:
\begin{equation}
H=H^{(2)}+\Delta H.
\end{equation}
The matrix elements of $H$ between the states presented above are expressed in terms of the basic overlap integral $I_{J}^{(0)}$ and its k-th derivatives, defined by
\begin{equation}
I_{J}^{(0)}(d^2)=\int_{0}^{1}P_{J}(y)e^{d^{2}P_{2}(y)}dy,\,\,\,I_{J}^{(k)}(x)=\frac{d^{k}I_{J}^{(0)}}{dx^k},\,\,\,x=d^{2},
\end{equation}
where $P_{J}$ denotes the Legendre polynomial of rank $J$. 

\renewcommand{\theequation}{3.\arabic{equation}}
\section{The vibrational  and near vibrational regimes}
\label{sec:level 3}

\subsection{Energies}

As already mentioned, in Refs.\cite{RadGhBa,Rad00,RadStSa}, it was proved that the projected states  go to the first three
highest seniority states respectively, when the parameter $d$ goes to zero.
For a easier writing, let us denote:

\begin{eqnarray}
\varphi^{i,v}_{JM}&=&\lim_{d\rightarrow  0}\varphi^i_{JM}(d),\nonumber\\
H^v&=&\lim_{d\rightarrow  0}H.
\label{fivib}
\end{eqnarray}
According to ref.\cite{RadGhBa,Rad00,RadStSa}, the vibrational limits for the projected states are:
\begin{eqnarray}
\varphi^{g,v}_{JM}&=&|\frac{J}{2},\frac{J}{2},0,J,M\rangle,
\nonumber\\
\varphi^{\gamma,v}_{JM}&=&|\left[\frac{J+3}{2}\right],\left[\frac{J+3}{2}\right],0,J,M\rangle,\; [...]-\rm{integer\; part},
\nonumber\\
\varphi^{\beta,v}_{JM}&=&\left[1-\frac{6}{7}\frac{J(2J+3)}{(J+7)(3J+10)}\right]^
{\frac{1}{2}}|\frac{J}{2}+3,\frac{J}{2}+3,1,J,M\rangle \nonumber\\
&+&\left[\frac{6}{7}\frac{J(2J+3)}{(J+7)(3J+10)}\right]^{\frac{1}{2}}|\frac{J}{2}+3,\frac{J}{2}+1,0,J,M\rangle
\nonumber\\
& \equiv &\varphi^{\beta v,1}_{JM}+\varphi^{\beta v,2}_{JM}.
\label{fivibggabe}
\end{eqnarray}
where, the standard notations for the states $|N,v,\alpha,J,M\rangle$, labeled by the number of bosons ($N$), seniority ($v$), missing quantum number ($\alpha$), angular momentum ($J$) and its projection on z axis ($M$), are used.These quantum numbers, except  $\alpha$, are given by the Casimir operators of the groups in the chain
$SU(5)\supset R(5)\supset R(3)\supset R(2)$. A complete description of these states may be found in Refs. \cite{Gh,RaGh}.
The vibrational limits  are related by the following equations:
\begin{eqnarray}
\varphi ^{\beta v,1}_{JM}&=&\left[\frac{3}{5}(3J+10)\right]^{-\frac{1}{2}}(b^{\dag}
b^{\dag}b^{\dag})_0\varphi^{g,v}_{JM},
\nonumber\\
\varphi ^{\beta v,2}_{JM}&=&\left[\frac{15}{7}\frac{J(2J+3)}{(J+7)^2(3J+10)}\right]^{\frac{1}{2}}
(b^{\dag}b^{\dag})_0\varphi^{\gamma,v}_{JM}.
\label{fibevib}
\end{eqnarray}
The vibrational limit for the band  energies are:
\begin{eqnarray}
E^{g,v}_J&=&11A_1J+A_2J(J+1),
\nonumber\\
E^{\gamma,v}&=&22A_1\left[\frac{J+3}{2}\right]+A_2J(J+1),
\nonumber\\
E^{\beta, v}_J&=&A_1\left[11(J+6)+\frac{12}{7}\frac{J(2J+3)}{3J+10}\right]\nonumber\\
&+&
A_2J(J+1)+\frac{3}{5}(3J+10)A_3.
\label{enevibra}
\end{eqnarray}
where [..] denotes the integer part.
Since the matrix elements of the model Hamiltonian between  states of ground and gamma bands
are vanishing in the vibrational limit, it results that the vibrational states are
eigenstates of $H$ in the restricted collective space. Moreover, one can prove
that this is
true in the whole boson space for ground and gamma band states of any angular
momentum. Concerning the beta band states, this property holds only for
the $J=0$ state. However, if one ignores the  component $\varphi ^{\beta v,2}_{JM}$ of the vibrational beta
states, the remaining component, i.e. $\varphi ^{\beta v,1}_{JM}$,  is an eigenstate of the
vibrational Hamiltonian:
\begin{eqnarray}
H^v\varphi^{\beta v,1}_{JM}&=&\left[11A_1(J+6)+A_2J(J+1)\right.\nonumber\\
&&\left.+\frac{3}{5}(3J+10)A_3\right]\varphi^{\beta v,1}_{JM}.
\label{hasv}
\end{eqnarray}
For small values of the deformation parameter the exact energies can be expressed as a power series in $d$. As a result the excitation energies of the three bands are written as compact formulas  depending on powers of $J(J+1)$ which are easy to be handled in numerical calculations.
As we have already  mentioned the matrix elements of the model Hamiltonian between the angular momentum projected states can be written as function of  the overlap integral $I_{J}^{(0)}$ and its k-th derivatives, $I_{J}^{(k)}$.
Taking into account the composition rule as well as the recurrence relations for the Legendre polynomial one can prove that the basic integrals satisfy the differential equation:

\begin{equation}
\frac{d^{2}I_{J}^{(0)}}{dx^{2}}-\frac{x-3}{2x}\frac{dI_{J}^{(0)}}{dx}-\frac{2x^{2}+J(J+1)}{4x^{2}}I_{J}^{(0)}=0,\,\,\,\,(x=d^{2}).
\label{difeq}
\end{equation}

The solution for this equation is:
\begin{equation}
I_{J}^{(0)}(d^2)=\frac{(J!)^{2}}{\left(\frac{J}{2}\right)!(2J+1)!}(6d^{2})^{\frac{J}{2}}e^{-\frac{d^{2}}{2}}{_{1}F_{1}}\left(\frac{1}{2}(J+1),J+\frac{3}{2};\frac{3}{2}d^{2}\right),
\end{equation}
where $_{1}F_{1}(a,b;z)$ is the hypergeometric function of the first kind.
The excitation energies of the ground, beta and gamma bands are functions of the the ratio $d^2\frac{I^{(1)}_{J}}{I^{(0)}_{J}}$ and its first three derivatives with respect to $d^2$. These quantities have the following vibrational limits:
\begin{eqnarray}
&&\lim_{d\rightarrow0}\left(d^{2}\frac{I_{J}^{(1)}}{I_{J}^{(0)}}\right)^{(k)}=\frac{1}{(2J+3)^{k}}\Bigg[\frac{J}{2}(\delta_{k,0}+\delta_{k,1})\nonumber\\
&&+9\frac{(J+1)(J+2)}{2J+5}\left(\delta_{k,2}+9\frac{\delta_{k,3}}{2J+7}\right)\Bigg],\nonumber\\
&&k=0,1,2,3.
\end{eqnarray}
These relations allows us to write down the Taylor expansion of $xI_{J}^{(1)}/I_{J}^{(0)}$ up to the third order in $x$ ($=d^2$):
\begin{eqnarray}
x\frac{I_{J}^{(1)}}{I_{J}^{(0)}}&=&\frac{J}{2}+\frac{J}{2(2J+3)}x+\frac{9}{2}\frac{(J+1)(J+2)}{(2J+3)^{2}(2J+5)}x^{2}\nonumber\\
&&+\frac{27}{2}\frac{(J+1)(J+2)}{(2J+3)^{3}(2J+5)(2J+7)}x^{3}.
\label{I1peI0}
\end{eqnarray}

Inserting the truncated power series of the $xI_{J}^{(1)}/I_{J}^{(0)}$,into the excitation energy expressions one obtains \cite{Raduta2}:
\begin{eqnarray}
E_{J}^{g}&=&22A_{1}\sum_{k=0}^{3}A_{J,k}^{(g)}x^{k}+A_{2}J(J+1)-\Delta E_{J},\\
E_{J}^{\gamma}&=&44A_{1}+\frac{A_{1}}{\sum_{k=0}^{3}Q_{J,k}^{(\gamma,0)}x^{k}}\left[\sum_{k=0}^{3}\left(22R_{J,k}^{(\gamma,0)}+5U_{J,k}^{(\gamma,0)}\right)x^{k}\right]\nonumber\\
&+&A_{2}J(J+1)+\Delta E_{J},\,J=even,\\
E_{J}^{\gamma}&=&44A_{1}+\frac{A_{1}}{\sum_{k=0}^{3}Q_{J,k}^{(\gamma,1)}x^{k}}\left[\sum_{k=0}^{3}\left(22R_{J,k}^{(\gamma,1)}+5U_{J,k}^{(\gamma,1)}\right)x^{k}\right]\nonumber\\
&+&A_{2}J(J+1),\,J=odd,\\
E_{J}^{\beta}&=&\frac{1}{\sum_{k=0}^{3}Q_{J,k}^{(\beta)}x^{k}}\Bigg\{A_{1}\sum_{k=0}^{3}\left(22R_{J,k}^{(\beta)}+5U_{J,k}^{(\beta)}\right)x^{k}
\nonumber\\
&+&\sum_{k=0}^{3}\left(A_{3}V_{J,k}^{(\beta)}+A_{4}dZ_{J,k}^{(\beta)}+A_{5}B_{J,k}^{(\beta)}\right)x^{k}\Bigg\}\nonumber\\
&+&A_{2}J(J+1).
\label{vibener}
\end{eqnarray}
The expansion coefficients $A, R, U, V, Z, B$ and the quantity $\Delta E$ are given in Appendix A.

\subsection{Reduced probability for E2 transitions}

CSM uses for the quadrupole transition operator the following expression:
\begin{eqnarray}
Q_{2\mu}&=&q_{h}(b_{\mu}^{\dagger}+(-)^{\mu}b_{-\mu})+q_{anh}((b^{\dagger}b^{\dagger})_{2\mu}+(bb)_{2\mu})\nonumber\\
         &\equiv&Q^h_{2\mu}+Q^{anh}_{2\mu}.
\end{eqnarray}
The anharmonic term is the lowest order term in bosons which brings a non-vanishing contribution to the E2 transition between a state from the beta band and a state from the ground band. 

Analytical expressions for transition probabilities are also possible. First we list the results for the limit of $d\to 0$ of the non-vanishing matrix elements of the terms involved in the transition operator.

The final results are \cite{Rad00,RadStSa}:
\begin{eqnarray}
\lim_{d\rightarrow 0}\langle\varphi^g_J||Q^h_2||\varphi^g_{J^{\prime}}\rangle&=&(1-
\delta_{J,J^{\prime}})
\left[\frac{2}{3}\frac{(J+J'+3)(J+J'+1)}{J+J'}\right]^{1/2}C^{J~2~J'}_{0~0~0}q_h,
\nonumber\\
\lim_{d\rightarrow 0}\langle\varphi^{\beta}_J||Q^h_2||\varphi^{\beta}_{J^{\prime}}\rangle
&=&(1-\delta_{J,J'})\left[\frac{2}{3}\frac{(J+J'+1)(J+J'+3)(3(J+J')+26)}{(J+J')(3(J+J')+14)}
\right]^{1/2}C^{J~2~J'}_{0~0~0}q_h,
\nonumber\\
\lim_{d\rightarrow 0}\langle \varphi^{\gamma}_J||Q^h_2||\varphi^{g}_{J}\rangle &=&
2\left[\frac{(J+1)(2J+3)}{3(J-1)(J+2)}\right]^{1/2}C^{J~~2~~J}_{2~-2~~0}q_h, J=\rm{even},
\nonumber\\
\lim_{d\rightarrow 0}\langle\varphi^{\gamma}_J||Q^h_2||\varphi^{g}_{J+1}\rangle&=&
-\left[\frac{6(J+1)(J+2)^2(2J+3)}{J(2J+1)(2J^2+5J+11)}\right]^{1/2}C^{J~~2~J+1}
_{2~-2~~0}q_h,~J=\rm{odd},
\nonumber\\
\lim_{d\rightarrow 0}\langle\varphi^{\beta}_J||Q^h_2||\varphi^{\gamma}_{J+2}\rangle&=&
2\left[\frac{6(2J+3)(2J+5)(2J+7)}{7(J+3)(J+4)(3J+10)}\right]^{1/2}C^{J~~2~J+2}_{0~~2~~2}q_h,
\nonumber\\
\lim_{d\rightarrow 0}\langle\varphi^{\beta}_J||Q^h_2||\varphi^{\gamma}_{J+1}\rangle&=&
-\left[\frac{108(J+2)(J+3)^2}{7(3J+10)(2J^2+9J+18)}\right]^{1/2}C^{J~~2~J+1}_{0~~2~~2}q_h,
\nonumber\\
\lim_{d\rightarrow 0}\langle\varphi^{\gamma}_J||Q^h_2||\varphi^{\gamma}_{J+2}\rangle&=&
\left[\frac{(J+1)(J+2)(2J+5)(2J+7)}{3(J-1)(J+3)(J+4)}\right]^{1/2}
C^{J~~2~J+2}_{2~~0~~2}q_h,
~J=\rm{even},
\nonumber\\
\lim_{d\rightarrow 0}\langle\varphi^{\gamma}_J||Q^h_2||\varphi^{\gamma}_{J+2}\rangle&=&
\left[\frac{(J+3)(2J+3)(4J^3+18J^2+45J+23)^2}{3J(2J+1)(J+1)(2J^2+13J+29)(2J^2+5J+11)}
\right]^{1/2}C^{J~~2~J+2}_{2~~0~~2}q_h,J=\rm{odd},
\nonumber\\
\lim_{d\rightarrow 0}\langle\varphi^{\gamma}_J||Q^h_2||\varphi^{\gamma}_{J+1}\rangle&=&
-\left[\frac{3(J+1)(J+2)^2(J+3)^2}{2(J-1)(2J+3)(2J^2+9J+18)}\right]^{1/2}C^{J~~2~J+1}_{2~~0~~2}  q_h,J=\rm{even},
\nonumber\\
\lim_{d\rightarrow 0}\langle\varphi^{g}_J||Q^{anh}_2||\varphi^{\beta}_{J-2}\rangle&=&
\left[\frac{4(J-1)J^2}{(2J-1)(2J+1)(3J+4)}\right]^{1/2}C^{J~~2~J-2}_{0~~0~~0}q_h.
\label{qredus}
\end{eqnarray}
Note that in the limit of large $J$, the  Alaga's rule \cite{Alaga} is valid even for the vibrational regime.

It is well known that the $B(E2)$ values are very sensitive to the small variation in both the wave functions and transition operator. Therefore we include in the expression of the matrix elements of the transition operator the first order Taylor expansion in terms of the deformation parameter $d$. Then the 
$B(E2)$ value characterizing a certain transition is obtained, in the Rose convention \cite{Rose}, by squaring the corresponding reduced matrix element. 

 Intraband transition matrix elements  are:
\begin{eqnarray}
\langle\phi_{J}^{g}||Q_{2}||\phi_{J-2}^{g}\rangle&=&\sqrt{\frac{J}{2}}\left(q_{h}-q_{anh}d\right),
\\
\langle\phi_{J}^{\beta}||Q_{2}||\phi_{J-2}^{\beta}\rangle&=&\sqrt{\frac{J(3J+10)}{2(3J+4)}}\left[q_{h}-\frac{q_{anh}d(3J+4)}{3J+10}\right],\\
\langle\phi_{J+2}^{\gamma}||Q_{2}||\phi_{J}^{\gamma}\rangle&=&\sqrt{\frac{J(2J+7)}{2(2J+3)}}\left[q_{h}-\frac{q_{anh}d(2J-13)}{2J+7}\sqrt{\frac{2}{7}}\right],\;\;J=even,\\
\langle\phi_{J}^{\gamma}||Q_{2}||\phi_{J-1}^{\gamma}\rangle&=&q_{h}d\sqrt{\frac{2(J-2)}{J(J-1)(2J+3)}}\Bigg[8-\frac{J-8}{2J-1}\\
&&+\frac{(J+2)(2J+3)}{(J+1)(2J+1)}+\frac{2(J-5)(J-1)(4J+3)}{(J+1)(2J-1)(2J+1)}\Bigg],\;\;J=even, \\
\langle\phi_{J+1}^{\gamma}||Q_{2}||\phi_{J}^{\gamma}\rangle&=&\sqrt{\frac{6(J+3)}{J(2J+3)}}\left[q_{h}-\frac{5q_{anh}d}{J+3}\sqrt{\frac{2}{7}}\right],\;\;J=even,\\
\langle\phi_{J+2}^{\gamma}||Q_{2}||\phi_{J}^{\gamma}\rangle&=&\sqrt{\frac{(J-1)(J+3)(J+4)}{2(J+1)(J+2)}}\left[q_{h}-q_{anh}d\frac{J+6}{J+4}\sqrt{\frac{2}{7}}\right], \;\;J=odd.
\end{eqnarray}

The interband transition matrix elements are:
\begin{eqnarray}
\langle\phi_{J}^{g}||Q_{2}||\phi_{J-2}^{\beta}\rangle&=&q_{anh}\sqrt{\frac{6J}{(3J+4)}}\left[1-\frac{3(34J^{2}+34J-29)}{14(2J-1)(2J+3)(3J+4)}d^{2}\right],\\
\langle\phi_{J}^{g}||Q_{2}||\phi_{J}^{\beta}\rangle&=&-2q_{anh}d\sqrt{\frac{3J(J+1)}{(2J-1)(2J+3)(3J+10)}},\\
\langle\phi_{J-2}^{g}||Q_{2}||\phi_{J}^{\beta}\rangle&=&q_{anh}d^{2}\frac{3(J-1)}{2J-1}\sqrt{\frac{6J}{(2J-3)(2J+1)(3J+10)}},\\
\langle\phi_{J}^{\gamma}||Q_{2}||\phi_{J}^{g}\rangle&=&
\sqrt{\frac{2(J+1)}{2J-1}}\nonumber\\
&&\times\left[q_{h}+\frac{2q_{anh}d(44J^{4}-210J^{3}-533J^{2}-15J+378)}{7(J-1)(J+1)(2J+3)^{2}}\sqrt{\frac{2}{7}}\right],\\
\langle\phi_{J}^{\gamma}||Q_{2}||\phi_{J-2}^{g}\rangle
&=&\sqrt{(2J+3)}\left[q_{anh}\sqrt{\frac{2}{7}}+\frac{3q_{h}d}{(2J+3)(2J-1)}\right],\\
\langle\phi_{J}^{\gamma}||Q_{2}||\phi_{J+2}^{g}\rangle
&=&\frac{6q_{h}d(J-1)}{(J+1)(2J+3)}\sqrt{\frac{J(J+2)(2J+5)}{(2J+1)(2J+3)}},\\
\langle\phi_{J-1}^{\gamma}||Q_{2}||\phi_{J}^{g}\rangle
&=&-\sqrt{\frac{(J-2)(2J+1)}{(J-1)(2J-1)}}\left[q_{h}+q_{anh}d\frac{(J+3)(J+4)}{(2J+1)(J+1)}\sqrt{\frac{2}{7}}\right],\\
\langle\phi_{J+1}^{\gamma}||Q_{2}||\phi_{J}^{g}\rangle
&=&-\sqrt{3(J+3)}\left[q_{anh}\sqrt{\frac{2}{7}}+\frac{q_{h}d}{(2J+3)}\right],
\end{eqnarray}
\begin{eqnarray}
\langle\phi_{J}^{\beta}||Q_{2}||\phi_{J+2}^{\gamma}\rangle
&=&\sqrt{\frac{6(2J+5)(2J+7)}{7(2J+1)(3J+10)}}\Bigg[q_{h}+q_{anh}d\frac{8J^{2}+42J+21}{(2J+3)(2J+7)}\sqrt{\frac{2}{7}}\Bigg],\\
\langle\phi_{J}^{\beta}||Q_{2}||\phi_{J}^{\gamma}\rangle
&=&4(J+5)\sqrt{\frac{3(J+1)}{7(2J-1)(3J+10)}}\nonumber\\
&&\times\left[q_{anh}\sqrt{\frac{2}{7}}+q_{h}d\frac{10J^{2}+13J-33}{2(J+1)(J+5)}\right],\\
\langle\phi_{J+2}^{\beta}||Q_{2}||\phi_{J}^{\gamma}\rangle&=&q_{anh}d\sqrt{\frac{3J(J+2)}{(2J+3)(3J+16)}}\frac{8(5J^{2}+17J-27)}{7(J+1)(2J+3)}.
\end{eqnarray}

In the limit $d\rightarrow0$ these expressions reproduce the m.e.  corresponding to vibrational case ($\ref{qredus}$), were  some transitions are forbidden \cite{Raduta3,Rad00}. Taking the next leading order of the transition m.e., the mentioned selection rules are washed out.

\renewcommand{\theequation}{4.\arabic{equation}}
\section{Large deformation regime}
\label{sec:level4}

One salient feature of CSM is the behavior of the projected states as function of the deformation parameter especially for the extreme limits of $d\to 0$ and  large $d$. While in the vibrational limit these are just multiphonon states in the rotational regime, i. e. for asymptotic values for deformation parameter $d$, the wave functions of the ground, beta  and gamma band states predicted by the liquid drop model\cite{Bohr} in the large deformation regime 
are nicely simulated. Indeed as proved in Ref.\cite{Raduta1}, writing the projected states in the intrinsic reference frame and then considering a large deformation $d$, one obtains:
\begin{eqnarray}
\varphi^i_{JM}&=&C_J\beta^{-1}e^{-(d-\frac{k\beta}{\sqrt{2}})^2 }
\left[\delta_{i,g}D^{J^*}_{M0}(\Omega_0)+
\delta_{i,\beta}\frac{4d^2}{9\sqrt{114}}
D^{J^*}_{M0}(\Omega_0)\right.\nonumber\\
&&\left.+\delta_{i,\gamma}\beta f_Jk\gamma (D^{J^*}_{M2}+(-)^JD^{J^*}_{M,-2},
(\Omega_0)\right],
\label{asimfi}
\end{eqnarray}
where $k$ is a constant defining the canonical transformation relating the quadrupole bosons and the quadrupole collective conjugate coordinates:
\begin{equation}
\alpha_{\mu}=\frac{1}{k\sqrt{2}}(b^{\dag}_{\mu}+(-)^{\mu}b_{-\mu}),~~
\pi_{\mu}=\frac{ik}{\sqrt{2}}((-)^{\mu}b^{\dag}_{-\mu}-b_{\mu}),
\label{alfa}
\end{equation}
while the constants $C_J$ and $f_J$ are

\begin{equation}
C_J=\frac{2}{3}\pi^{-\frac{1}{4}}k^{2/3}(2J+1)^{1/2},~f_J=-\sqrt{2}(8+(-)^{J+1})^{-1/2}.
\label{cj}
\end{equation}
It is worth noticing that the model Hamiltonian yields for the ground band similar excitation energies as the effective Hamiltonian
\begin{equation}
H_{eff}=11A_1\hat{N}+A_2\hat{J}^2.
\end{equation}
Averaging this Hamiltonian on a vibrational ground band state one obtains a quadratic expression in $N$, the number of bosons in the considered state:
\begin{equation}
\langle H_{eff}\rangle =N(11A_1+2A_2+4A_2N).
\end{equation}
In the asymptotic region for $d$ the average matrix element of $H_{eff}$ is \cite{RadGhBa} proportional to $J(J+1)$:
\begin{equation}
\langle H_{eff}\rangle =J(J+1)\left(\frac{11A_1}{6d^2}+A_2\right).
\end{equation}
For the intermediate situation for the deformation parameter $d$, we may use for energies either rational functions of $d$ with the coefficients being functions of the angular momentum as given in the previous section, or asymptotic expansion for the matrix elements in power of $1/x$. The later version was developed in Ref.\cite{Raduta3}. Here we sketch the ideas and give the final results.
\subsection{Energies}

The asymptotic expressions for the matrix elements are obtained by considering the behavior of the overlap integral $I^{(0)}_J$ for large $d$. This is obtained by using the asymptotic expression for the hypergeometric function:
 \begin{equation}
F(a,c;z)=\frac{\Gamma(c)}{\Gamma(a)}e^{z}z^{a-c}[1+\mathcal{O}(|z|^{-1})],
\end{equation}
One finds that the dominant term of the asymptotic form of $I_{J}^{(0)}$ is:
\begin{equation}
I_{J}^{(0)}\sim\frac{e^{x}}{3x}.
\label{firstappr}
\end{equation}
This suggests as trial function for the quantity $I_{J}^{(0)}$  satisfying the differential equation (\ref{difeq})  the following series:
\begin{equation}
I_{J}^{(0)}=e^{x}\sum_{n=1}A_{n}x^{-n}.
\label{solut}
\end{equation}
This series expansion together with the differential equation offer a recurrence relation for the series coefficients.
\begin{equation}
A_{n+1}=\frac{A_{n}}{6n}(2n+J)(2n-J-1).
\end{equation}
Using the asymptotic form (\ref{firstappr}) as the limit condition, which infers $A_{1}=\frac{1}{3}$, the solution (\ref{solut}) is completely determined.

For big values of the deformation parameter, the series can be approximated by a truncation, such that one arrive at the following expression
\begin{eqnarray}
x\frac{I_{J}^{(1)}}{I_{J}^{(0)}}&=&x-1-\frac{1}{3x}-\frac{5}{9x^{2}}-\frac{37}{27x^{3}}+\left(\frac{1}{6x}+\frac{5}{18x^{2}}+\frac{13}{18x^{3}}\right)J(J+1)\nonumber\\
&&-\frac{1}{54x^{3}}J^{2}(J+1)^{2}+\mathcal{O}(x^{-4}).
\label{truncser}
\end{eqnarray}

This approximation can be substantially improved. Indeed, let us write the differential equation (\ref{difeq}) in the form
\begin{equation}
x\left(x\frac{I_{J}^{(1)}}{I_{J}^{(0)}}\right)'+\left(x\frac{I_{J}^{(1)}}{I_{J}^{(0)}}\right)^{2}-\frac{x-1}{2}\left(x\frac{I_{J}^{(1)}}{I_{J}^{(0)}}\right)-\frac{2x^{2}+J(J+1)}{4}=0
\end{equation}
and  replace the first term  by the derivative of the expression (\ref{truncser}). Obviously,  one obtains a quadratic equation for the quantity $xI_{J}^{(1)}/I_{J}^{(0)}$ whose positive solution is:
\begin{equation}
x\frac{I_{J}^{(1)}}{I_{J}^{(0)}}=\frac{1}{2}\left[\frac{x-2}{2}+\sqrt{G_{J}}\right],
\end{equation}
where
\begin{eqnarray}
G_{J}&=&\frac{9}{4}x(x-2)+\left(J+\frac{1}{2}\right)^{2}-\frac{4}{9x}\left(3+\frac{10}{x}+\frac{37}{x^{2}}\right)\nonumber\\
&&+\frac{2}{3x}\left(1+\frac{10}{3x}+\frac{13}{x^{2}}\right)J(J+1)-\frac{2J^{2}}{9x^{3}}(J+1)^{2}.
\end{eqnarray}

Note  that the mixing m.e. between ground and $\gamma$ states  are negligible within the approximation of large deformation.

Using the approximation (\ref{truncser}), the energies of the $\beta$ and $\gamma$ bands can be written as follows:
\begin{eqnarray}
E_{J}^{\beta}&=&\frac{1}{P_{J}^{\beta}}\left[A_{1}S_{J}^{\beta}+A_{3}F_{J}^{\beta}\right]+A_{2}J(J+1),\\
E_{J}^{\gamma}&=&A_{1}\frac{S_{J}^{\gamma}}{P_{J}^{\gamma}}+A_{2}J(J+1),
\end{eqnarray}
The polynomials $P,S,F$, in $J(J+1)$,  are given in Appendix B.
To these equations we add the equations determining the excitation energies in the ground band:
\begin{equation}
E_{J}^g=11A_1\left[\frac{x-2}{2}+\sqrt{G}\right]+A_2J(J+1).
\end{equation}

In order to obtain a good agreement for $\beta$-band energies the use of an additional term accompanied by the $A_{4}$  or $A_5$ parameter is necessary for some few cases. For these additional terms the following asymptotic relations are used:
\begin{eqnarray}
\langle\phi_{JM}^{\beta}|\Omega_{\beta}^{\dagger}\Omega_{\beta'}^{2}+h.c.|\phi_{JM}^{\beta}\rangle
&=&\frac{96d}{5\sqrt{70}}\Bigg(\frac{x}{2}-\frac{T_{J}^{4,\beta}}{P_{J}^{\beta}}\Bigg),\nonumber\\
\langle\phi_{JM}^{\beta}|\Omega_{\beta'}^{\dagger2}\Omega_{\beta'}^{2}|\phi_{JM}^{\beta}\rangle&=&\frac{32}{875}\frac{T^{5,\beta}}{P_{J}^{\beta}},
\end{eqnarray}
with the factors $T_J^{n,\beta}$, with n=4,5, and  $P_{J}^{\beta}$ defined in Appendix B.

\subsection{Reduced probabilities for the E2 transitions}
Taking the asymptotic limit of the exact m.e. of the quadrupole operator, one obtains very simple formula for transition m.e. for large deformation case. The asymptotic expressions for the reduced m.e. of the harmonic quadrupole transition operator are \cite{Raduta3}:
\begin{eqnarray}
\langle\phi_{J}^{i}||Q_{2}^{h}||\phi_{J'}^{i}\rangle &=&2dq_{h}C^{J\,2\,J'}_{K_{i}0\,K_{i}},\,\,\,i=g,\beta,\gamma,\,\,\,K_{i}=-2\delta_{i\gamma},\\
\langle\phi_{J}^{\gamma}||Q_{2}^{h}||\phi_{J'}^{g}\rangle &=&\sqrt{2}q_{h}C^{\,J\,2\,J'}_{-2\,2\,0},\\
\langle\phi_{J}^{\beta}||Q_{2}^{h}||\phi_{J'}^{\gamma}\rangle&=&\frac{2}{3\sqrt{19}}q_{h}C^{J\,2\,J'}_{0\,-2\,-2},
\label{Alag1}
\end{eqnarray}
while the $\beta$ and ground band states are connected by anharmonic part of $Q_{2\mu}$:
\begin{equation}
\langle\phi_{J}^{\beta}||Q_{2}^{anh}||\phi_{J'}^{g}\rangle =2\sqrt{\frac{7}{19}}q_{anh}C^{J\,2\,J'}_{0\,0\,0}.
\label{Alag2}
\end{equation}
Note that in the asymptotic limit of the deformation parameter $d$, the projected functions are similar to that of the liquid drop model in the strong coupling regime. The Clebsch-Gordan factorization of the transition probabilities is known in literature as Alaga's rule \cite{Alaga}.
Thus, we may say that our description of the deformed nuclei is consistent with the Alaga's rule.

\renewcommand{\theequation}{5.\arabic{equation}}
\section{Numerical results}
\label{sec:level5}
The analytical expressions for energies and transition probabilities presented in the previous sections were applied to 42 nuclei from which 18 are considered to be near vibrational while 24 are well deformed. The results are compared with the available data for both energies and reduced transition probabilities. Results are presented in a slightly different order than in the previous sections. Indeed, we divide this section into two parts one devoted to energies and one to  e.m. transitions. The reason is that we aim at pointing out the change in the spectrum structure  and separately in the behavior of the transition probabilities when one passes from a near vibrational to a deformed regime. 

\subsection{Energies}
Energies for near vibrational nuclei have been calculated with Eq.(3.10)-(3.13). The parameters involved were calculated by a least square procedure.
The results are listed in Table I. Therein we also give the root mean square for the  deviations of the calculated excitation energies from the corresponding experimental data, denoted by $\chi$, the total number of states in the three bands, the ratio $E_{4_1^+}/ E_{2_1^+}$ and the nuclear deformation $\beta_2$. The mentioned ratio indicates how far are we from the vibrational limit which is 2. Another measure of this departure is of course the deformation parameter $d$. For nuclei close to a spherical shape, $d$ is under-unity, while for a transitional nucleus $d$ may become larger than unity. Since the energies are power series of $x (=d^2)$ it is necessary to comment on the convergence of such series. A detailed study of this issue was presented in Ref.\cite{RaBuFa},
where it was shown that the convergence radius of the series associated to the overlap integral $I^{(0)}_J$ si larger than unity. As a matter of fact this property allows us to  consider the nuclei from Table I with $d$ larger than unity but smaller than the convergence radius found in Ref.\cite{RaBuFa}
as belonging to the class of near vibrational isotopes. 

Excitation energies in ground, beta and gamma bands are presented in Fig. 1 as function of angular momentum. The case of $^{152}$Gd is included in Fig. 2 where the other even isotopes of Gd are presented. From Fig. 1 we notice that $^{188,190,192}$Os and $^{190,194,196}$Pt the three bands are well separated and evolve almost parallel with each other. All the mentioned nuclei are gamma unstable since the band gamma is less excited than the band beta. In $^{102}$Pd and $^{126}$Xe the excited bands cross each other and they become gamma stable after the crossing point. In $^{182}$Pt and $^{186}$Pt
the excited bands are close to each other, this feature being associated with the $SU(3)$ symmetry. We remark that in $^{154}$Dy the excited bands and ground band are close to each other, which reflects the existence of a very flat potential in the $\beta$ and very flat potential in the $\gamma$ variable. A peculiar structure of the three bands is seen for $^{186}$Hg where the beta band crosses the ground  band becoming yrast state from $J=4$.
As shown in Fig.2, $^{152}$Gd is a gamma stable nucleus.  
\begin{table}[htbp!]
\caption{The fitted parameters, $d, A_1, A_2, A_3, A_4, A_5$  determining the ground, $\gamma$ and $\beta$ energies in the limit of small $d$, i.e. near vibrational regime, for 13 nuclei.  Also we give the r.m.s. for  the deviations of the calculated and experimental energies, denoted by $\chi$, the total number of states in the considered three bands, the ratio $E_{4_1^+}/ E_{2_1^+}$ and the nuclear deformation $\beta_2$ \cite{Lala}. In the fist column are listed the nuclei and the reference from where the experimental data are taken.
}
\begin{center}
\scriptsize{
\begin{tabular}{|l|c|r|r|r|r|r|r|r|c|}
\hline
Nucleus&$E_{4^{+}_{1}}/E_{2^{+}_{1}}$&$d$~~~~~~&$A_{1}$ [keV]&$A_{2}$ [keV]&$A_{3}$ [keV]&$A_{4}$ [keV]&$A_{5}$ [keV]&$\chi$ [keV]&Number of\\
&&&&&&&&&states\\
\hline
$^{102}$Pd\cite{DeFrenne}&2.293&1.45899&31.38296&8.58408&-48.05379&0~~~~~~~~&0~~~~~~~~&29.29906&11\\
$^{126}$Xe\cite{Katakura}&2.424&0.67610&16.98803&13.13610&-90.02417&-93.79735&0~~~~~~~~&37.42684&15\\
$^{152}$Gd\cite{Artna1}  &2.194&1.51491&22.18661&2.18499&47.49858&77.67379&34.40286&33.04687&20\\
$^{154}$Dy\cite{Reich1}  &2.233&1.53241&20.91108&3.36512&15.08504&51.90471&133.94467&44.59838&30\\
$^{188}$Os\cite{Balraj1} &3.083&1.62319&12.52609&9.63624&0~~~~~~~~&-9.81638&0~~~~~~~~&27.74638&13\\
$^{190}$Os\cite{Balraj2} &2.934&1.59990&11.27512&12.45678&0~~~~~~~~&0~~~~~~~~&163.08491&11.93490&14\\
$^{192}$Os\cite{Baglin1} &2.820&1.61011&10.53492&13.70670&0~~~~~~~~&0~~~~~~~~&263.51135&40.11629&16\\
$^{182}$Pt\cite{Balraj3} &2.705&1.69634&17.19040&3.67131&115.45535&99.68881&-11.03489&38.35839&24\\
$^{186}$Pt\cite{Baglin2} &2.560&1.67744&17.34034&4.29899&109.10067&98.17453&-17.85114&35.31432&18\\
$^{190}$Pt\cite{Balraj2} &2.492&0.74815&12.23722&11.16625&0~~~~~~~~&-3.45036&0~~~~~~~~&54.24460&12\\
$^{194}$Pt\cite{Browne1} &2.470&0.83137&11.49716&15.25325&-89.44485&-109.08323&0~~~~~~~~&27.23740&11\\
$^{196}$Pt\cite{Chun}    &2.465&0.95083&13.08631&15.24711&0~~~~~~~~&-11.04808&0~~~~~~~~&57.96742&13\\
$^{186}$Hg\cite{Baglin2} &2.665&0.92388&20.30171&4.24561&-138.41747&-16.77953&1063.07400&53.10842&24\\
\hline
\end{tabular}}
\end{center}
\end{table}

\begin{figure}[htbp!]
\begin{center}
\includegraphics[width=0.98\textwidth]{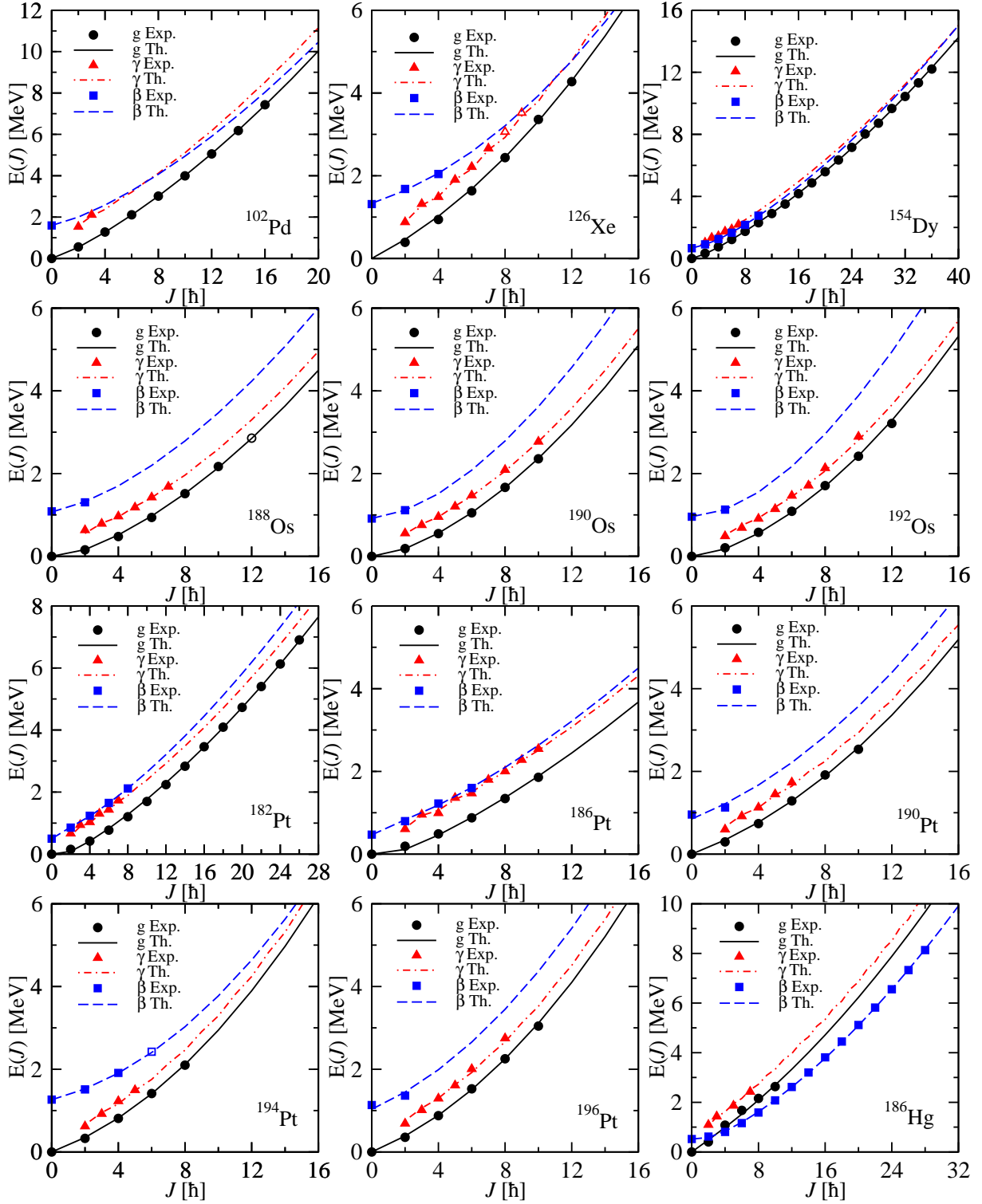}
\end{center}
\vspace{-0.8cm}
\caption{\footnotesize Energy spectra of ground, $\gamma$ and $\beta$ bands described by means of vibrational formulas for nuclei belonging to different nuclear phases. Open symbols denote uncertain or with possible band assignment experimental points, which were not taken into account in the fitting procedure. Experimental data are taken from 
\cite{DeFrenne,Katakura,Reich1,Balraj1,Balraj2,Baglin1,Balraj3,Baglin2,Browne1,Chun}.}
\end{figure}
\label{Fig. 1}
\clearpage

The results concerning the fitted parameters for the well deformed nuclei are given in Table II, Table III and Table IV. Results for Gd isotopes are given separately in Table II and Fig. 2. Except $^{154}$Gd which seems to be the critical nucleus in the phase transition from SU(5) to SU(3) symmetry
\cite{RaFa}, all isotopes from Table II are characterized by values of $d$ close to the rotational limit which is 3.3. From Fig. 2 one sees that the first three isotopes exhibit the features of a gamma stable nucleus while  the heaviest two isotopes are gamma unstable nuclei. In $^{158}$Gd, the excited bands have the states of even angular momentum degenerate which results in exhibiting a SU(3) symmetry. For high odd angular momenta in gamma band of $^{154}$Gd and $^{156}$Gd the moment of inertia becomes different from that of even angular momentum states. This is caused by the series truncation, which does not assures the expansion convergence in this particular region of $J$.

\begin{table}[htbp!]
\caption{The same as in Table I but for the Gd isotopic chain.}
\begin{center}
\scriptsize{
\begin{tabular}{|l|c|r|r|r|r|r|c|}
\hline
Nucleus&$E_{4^{+}}/E_{2^{+}}$&$d$~~~~~~&$A_{1}$ [keV]&$A_{2}$ [keV]&$A_{3}$ [keV]&$\chi$ [keV]&Number of\\
&&&&&&&states\\
\hline
$^{154}$Gd\cite{Reich1} &3.015&2.72583&18.68743&5.43251&-18.89754&38.51726&25\\
$^{156}$Gd\cite{Reich2} &3.239&3.08725&22.03879&3.84937&-15.81253&31.48198&35\\
$^{158}$Gd\cite{Helmer1}&3.288&3.30765&21.45168&4.67653&-10.95906&10.07559&15\\
$^{160}$Gd\cite{Reich3} &3.302&3.31382&17.40521&5.69801&-1.72684&7.01074&19\\
$^{162}$Gd\cite{Helmer2}&3.291&3.28976&15.19002&5.83600&2.43502&1.04853&11\\
\hline
\end{tabular}}
\end{center}
\end{table}

\begin{figure}[htbp!]
\begin{center}
\includegraphics[trim = 0mm 0mm 0mm 0mm,clip,width=0.98\textwidth]{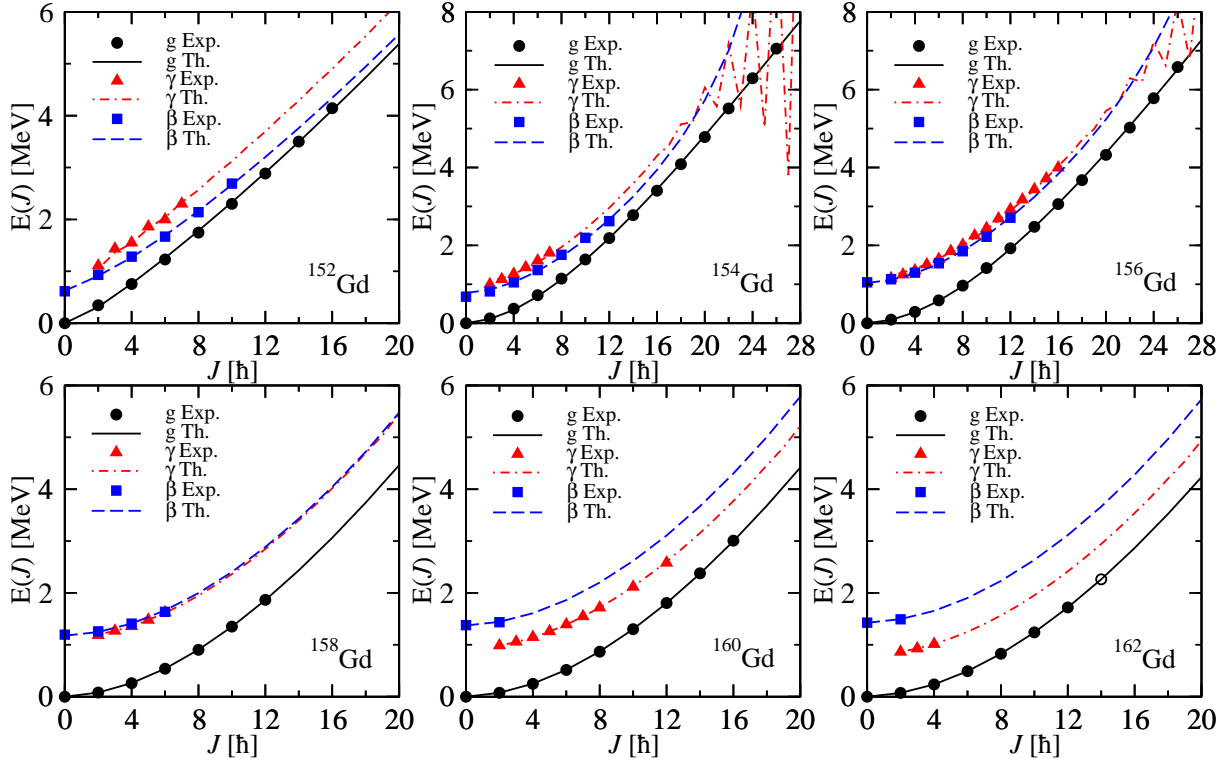}
\end{center}
\vspace{-0.8cm}
\caption{\footnotesize The isotopic chain of Gd. First nucleus is treated like a near vibrational one, later ones are deformed nuclei described by means of asymptotic regime formulas with 4 parameters. Open symbols denote uncertain or with possible band assignment experimental points, which were not taken into account in the fitting procedure. Experimental data are taken from \cite{Artna1,Reich1,Reich2,Helmer1,Reich3,Helmer2}.}
\label{Fig. 2}
\end{figure}

\clearpage

Fitted parameters for some transuranic nuclei are given in Table III, while the calculated excitation energies are compared with the available corresponding data in Fig.3. Except for $^{228,230}$Th, which exhibit a triaxial shape \cite{Rabuga}, the listed nuclei have a ratio $E_{4_1^+}/ E_{2_1^+}$
close to the rotational limit. From Table III one remarks the high accuracy for the theoretical description. Except for $^{128}$Th,$^{232}$U and $^{240}$Pu where
the two excited bands are only slightly split apart, for other nuclei the excited states  relative position reclaim on ideal SU(3) symmetry.

\begin{table}[htbp!]
\caption{The same as in Table I but for other nuclei.}
\begin{center}
\scriptsize{
\begin{tabular}{|l|c|r|r|r|r|r|c|}
\hline
Nucleus&$E_{4^{+}}/E_{2^{+}}$&$d$~~~~~~&$A_{1}$ [keV]&$A_{2}$ [keV]&$A_{3}$ [keV]&$\chi$ [keV]&Number of\\
&&&&&&&states\\
\hline
$^{228}$Th\cite{Artna2,Groger}&3.235&3.20609&17.94631&1.58239&-7.03770&8.61377&21\\
$^{230}$Th\cite{Akovali1}     &3.273&3.21904&13.85032&2.82743&-9.72828&4.42896&20\\
$^{232}$Th\cite{Schmorak1}    &3.284&3.37319&14.20081&2.53605&-7.87469&22.76356&40\\
$^{232}$U\cite{Schmorak1}     &3.291&3.44137&15.73529&2.17430&-10.14916&5.28270&19\\
$^{234}$U\cite{Akovali2}      &3.296&3.66457&17.08527&1.73226&-9.05098&8.89688&25\\
$^{236}$U\cite{Schmorak2}     &3.304&3.61576&17.51131&1.92143&-8.44386&2.28183&22\\
$^{238}$U\cite{Chukreev1}     &3.303&3.74042&19.38859&1.60137&-10.11735&26.83668&44\\
$^{238}$Pu\cite{Chukreev1}    &3.311&3.96825&18.78524&2.38340&-8.40586&1.03977&18\\
$^{240}$Pu\cite{Chukreev2}    &3.309&4.06356&20.88751&1.83394&-11.17675&3.34091&19\\
$^{242}$Pu\cite{Akovali3}     &3.307&3.87399&20.15834&1.86784&-10.08071&2.90317&16\\
$^{246}$Cm\cite{Akovali4}     &3.314&4.17501&20.61062&2.28038&-7.00481&1.34909&9\\
$^{248}$Cm\cite{derMat}       &3.309&3.78738&19.18411&1.77579&-7.55677&4.11945&19\\
\hline
\end{tabular}}
\end{center}
\end{table}

\begin{figure}[htbp!]
\begin{center}
\includegraphics[width=0.98\textwidth]{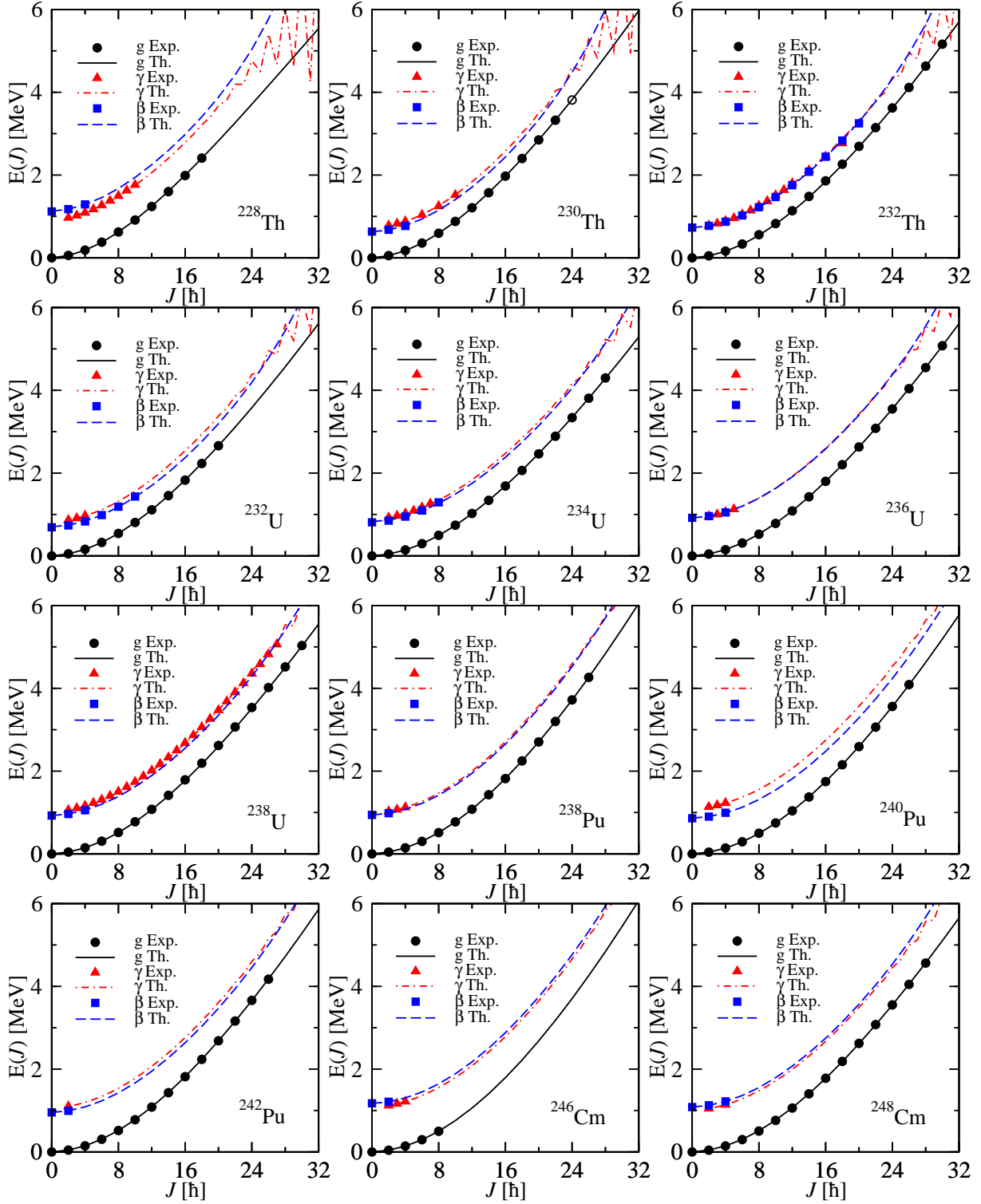}
\end{center}
\vspace{-0.8cm}
\caption{\footnotesize Energy spectra of ground, $\gamma$ and $\beta$ bands described by means of asymptotic formulas with 4 parameters for $SU(3)$ nuclei from transuranic region. $^{228}$Th and $^{230}$Th are possible candidates for triaxial nuclei. Open symbols denote uncertain or with possible band assignment experimental points, which were not taken into account in the fitting procedure. Experimental data are taken form \cite{Artna2,Groger,Akovali1,Schmorak1,Akovali2,Schmorak2,Chukreev1,Chukreev2,Akovali3,Akovali4}.}
\label{Fig. 3}
\end{figure}
\clearpage

In Table IV the fitted parameters for some deformed rare earth nuclei are presented.
The energies ratio of the ground band states $4^+$ and $2^+$ ranges from 2.9 to 3.3.
The lowest values 2.929, 3.009, 3.022 suggest that the nuclei to which they are assigned 
$^{150}$Nd, $^{152}$Sm and $^{178}$Os satisfy the X(5) symmetry. 
Four nuclei, $^{176}$Hf, $^{182,186}$W and $^{186}$Os, have the signature of triaxial nuclei. The remaining nuclei have the above mentioned ratio close to 3.3, i.e. they belong to the rotational
behaving nuclei. The plots from Fig. 4 show that for some nuclei like $^{152}$Sm, $^{172}$Yb and $^{186}$W, the excited bands do not intersect each other. In $^{162}$Dy the intersection is associated to the transition from the gamma unstable to the gamma stable behavior, while in 
$^{150}$Nd the transition is, by contrary, from gamma stable to gamma unstable regime. For describing the complex structure of the three bands the five parameters formulas are used. Exception is for $^{164}$Dy where  the set of three parameters formulas is used. The reduced number of the 
necessary parameters is explained by the fact that  here the beta band is missing.

\begin{table}[htbp!]
\caption{The same as in Table I but  for strongly deformed nuclei from the rare earth region.}
\begin{center}
\scriptsize{
\begin{tabular}{|l|c|r|r|r|r|r|r|c|}
\hline
Nucleus&$E_{4^{+}}/E_{2^{+}}$&$d$~~~~~~&$A_{1}$ [keV]&$A_{2}$ [keV]&$A_{3}$ [keV]&$A_{4}$ [keV]&$\chi$ [keV]&Number of\\
&&&&&&&&states\\
\hline
$^{150}$Nd\cite{derMat}&2.929&2.56790&19.49601&3.98450&-317.74302&-218.00288&21.85874&14\\
$^{152}$Sm\cite{Artna1}&3.009&2.69296&21.30931&4.03488&-20.22711&4.04777&47.24281&24\\
$^{162}$Dy\cite{Helmer3}&3.294&3.09941&15.51332&6.30447&157.11712&113.15065&14.18735&28\\
$^{164}$Dy\cite{Balraj4}&3.301&3.05374&13.15152&6.08637&0.0&0.0&4.60156&20\\
$^{166}$Er\cite{Shursh}&3.289&2.83539&13.80805&5.81240&89.40226&60.78307&5.90918&23\\
$^{172}$Yb\cite{Balraj5}&3.305&3.69655&26.79492&4.83027&34.86006&36.41885&24.74324&20\\
$^{174}$Yb\cite{Browne2}&3.310&3.78841&29.85375&4.05312&-8.45732&4.17396&2.82653&17\\
$^{176}$Hf\cite{Browne3}&3.284&3.20357&25.26813&4.06412&33.53111&37.14456&28.12626&22\\
$^{182}$W\cite{Balraj3}&3.291&3.20632&21.88739&7.37645&-125.28004&-81.67164&9.13062&15\\
$^{186}$W\cite{Baglin2}&3.234&2.58620&12.34463&11.44538&-79.27311&-56.40554&15.05515&18\\
$^{178}$Os\cite{Browne4}&3.022&2.45573&16.62605&5.68882&-189.91058&-125.86003&37.91601&16\\
$^{186}$Os\cite{Baglin2}&3.165&2.37861&13.52361&10.41550&121.91976&91.28338&32.58668&25\\
\hline
\end{tabular}}
\end{center}
\end{table}

\begin{figure}[htbp!]
\begin{center}
\includegraphics[width=0.98\textwidth]{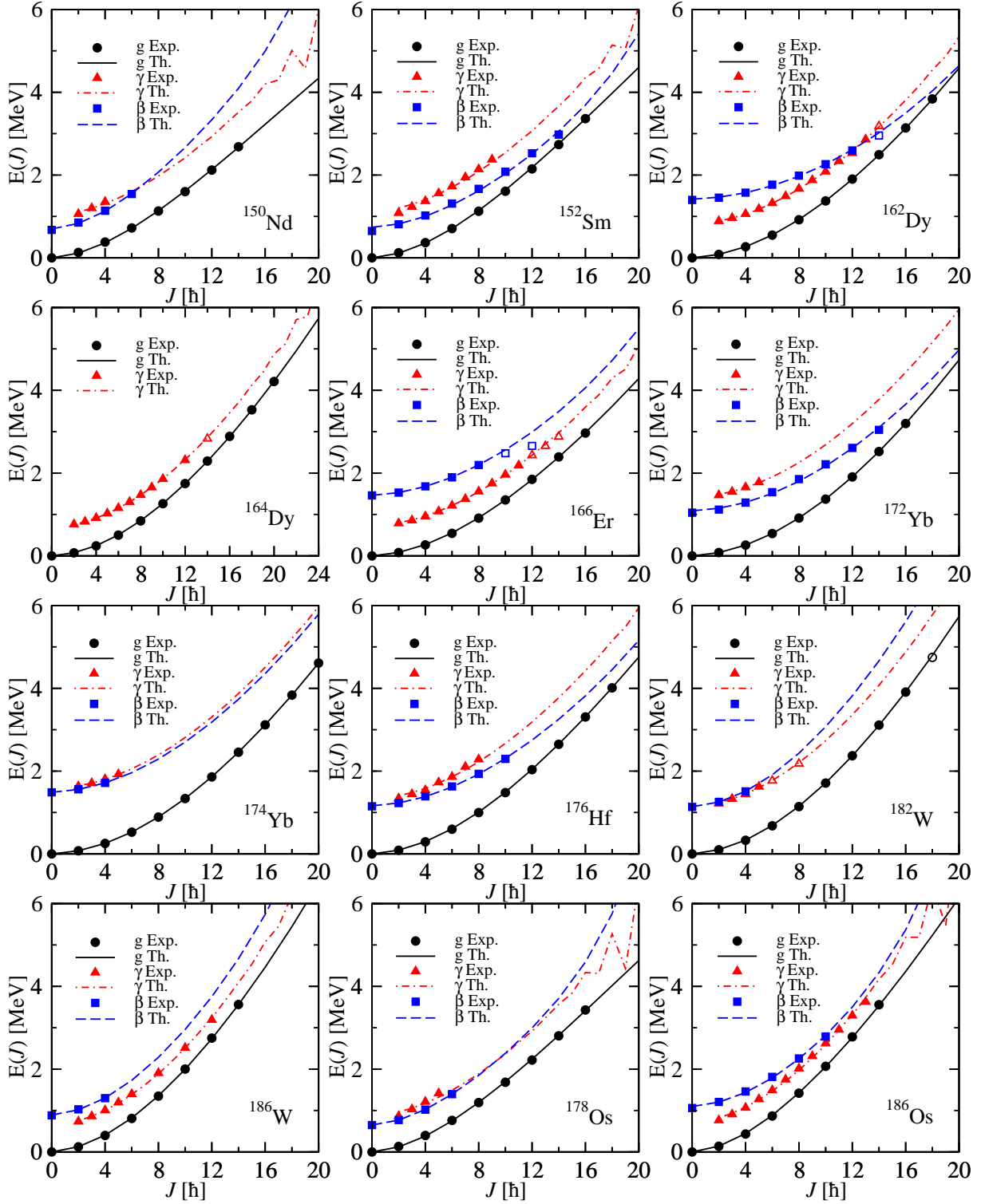}
\end{center}
\vspace{-0.8cm}
\caption{\footnotesize Energy spectra of ground, $\gamma$ and $\beta$ bands described by means of asymptotic formulas with 5 parameters for strongly deformed nuclei from the rare earth region. Open symbols denote uncertain experimental points or probable band assignment, and were not taken into account in the fitting procedure. Experimental data are taken from \cite{Balraj3,Baglin2,derMat,Artna1,Helmer3,Balraj4,Shursh,Balraj5,Browne2,Browne3,Browne4}.}
\end{figure}
\label{Fig. 4}
\clearpage

\begin{figure}[htbp!]
\begin{center}
\includegraphics[width=0.98\textwidth]{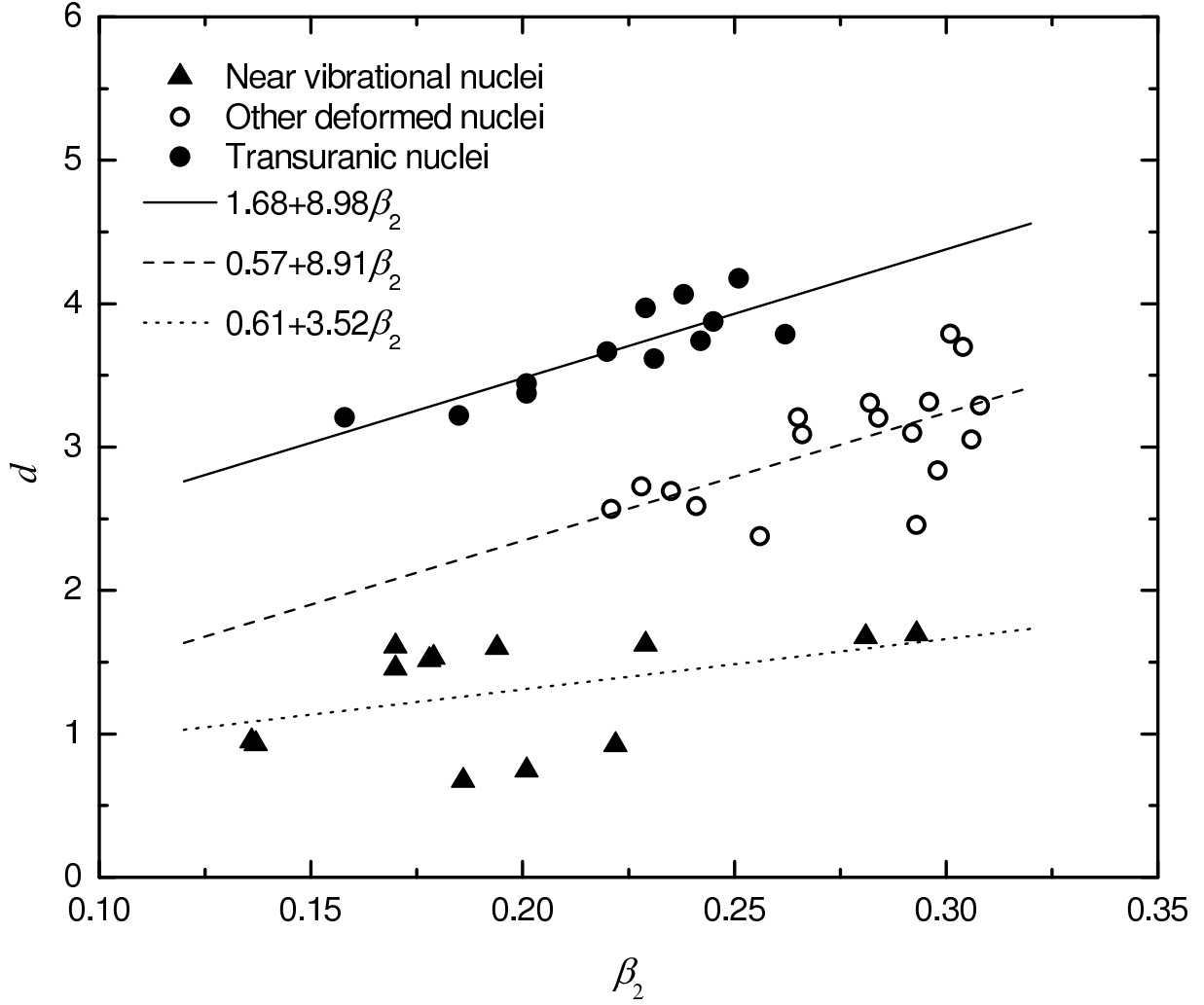}
\end{center}
\caption{The deformation parameter $d$ as function of the nuclear deformation $\beta_2$ taken from Ref.\cite{Lala}.}
\label{Fig. 5}
\end{figure}

\begin{figure}[htbp!]
\begin{center}
\includegraphics[width=0.98\textwidth]{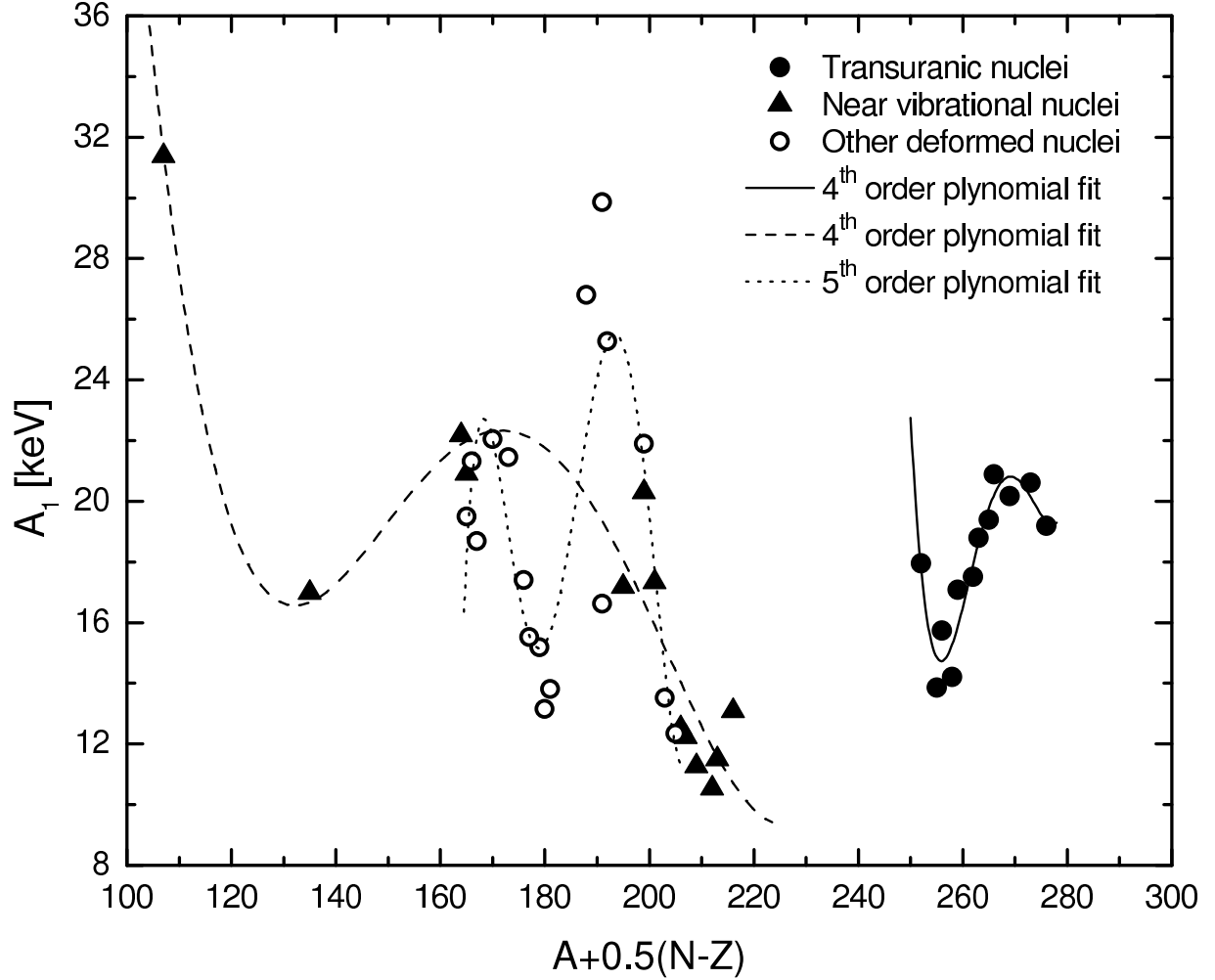}
\end{center}
\caption{The parameter $A_1$ as yielded by the fitting procedure as function of A-(N-Z)/2. The points corresponding to each group of nuclei were interpolated by a suitable polynomial.}
\label{Fig. 6}
\end{figure}

\begin{figure}[htbp!]
\begin{center}
\includegraphics[width=0.98\textwidth]{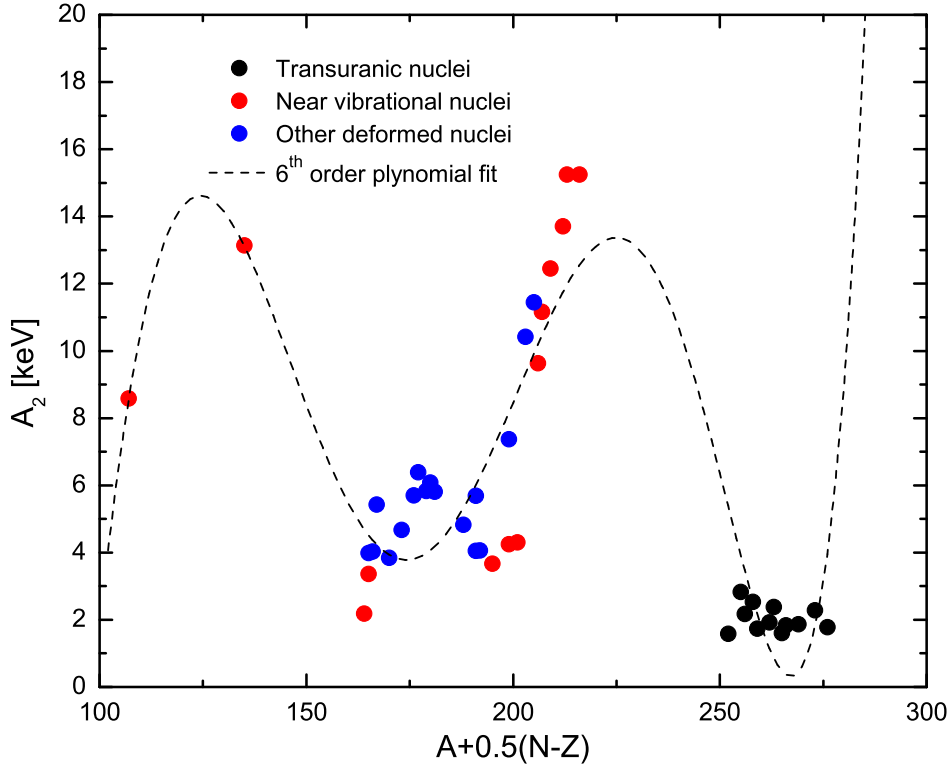}
\end{center}
\caption{The same as in Fig. 6, but for $A_2$.}
\label{Fig. 7}
\end{figure}
\begin{figure}[htbp!]
\begin{center}
\includegraphics[width=0.98\textwidth]{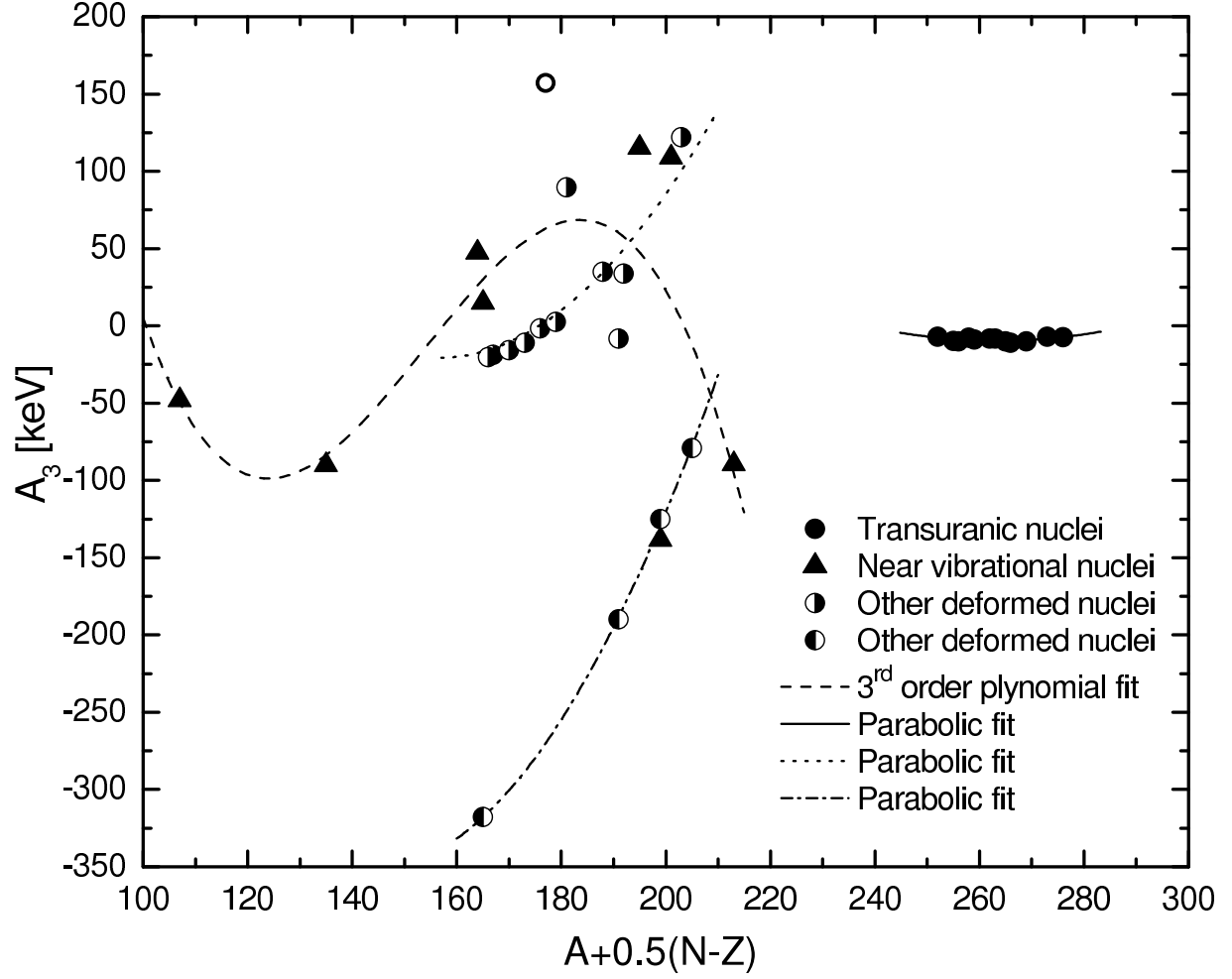}
\end{center}
\caption{The parameter $A_3$ as yielded by the fitting procedure as function of $A+(N-Z)/2$. There are four groups of nuclei, three of them having $A_3$ which could be interpolated by distinct parabolas and one with  the corresponding $A_3$ parameters interpolated by a third order polynomial.}
\label{Fig. 8}
\end{figure}
\begin{figure}[htbp!]
\begin{center}
\includegraphics[width=0.98\textwidth]{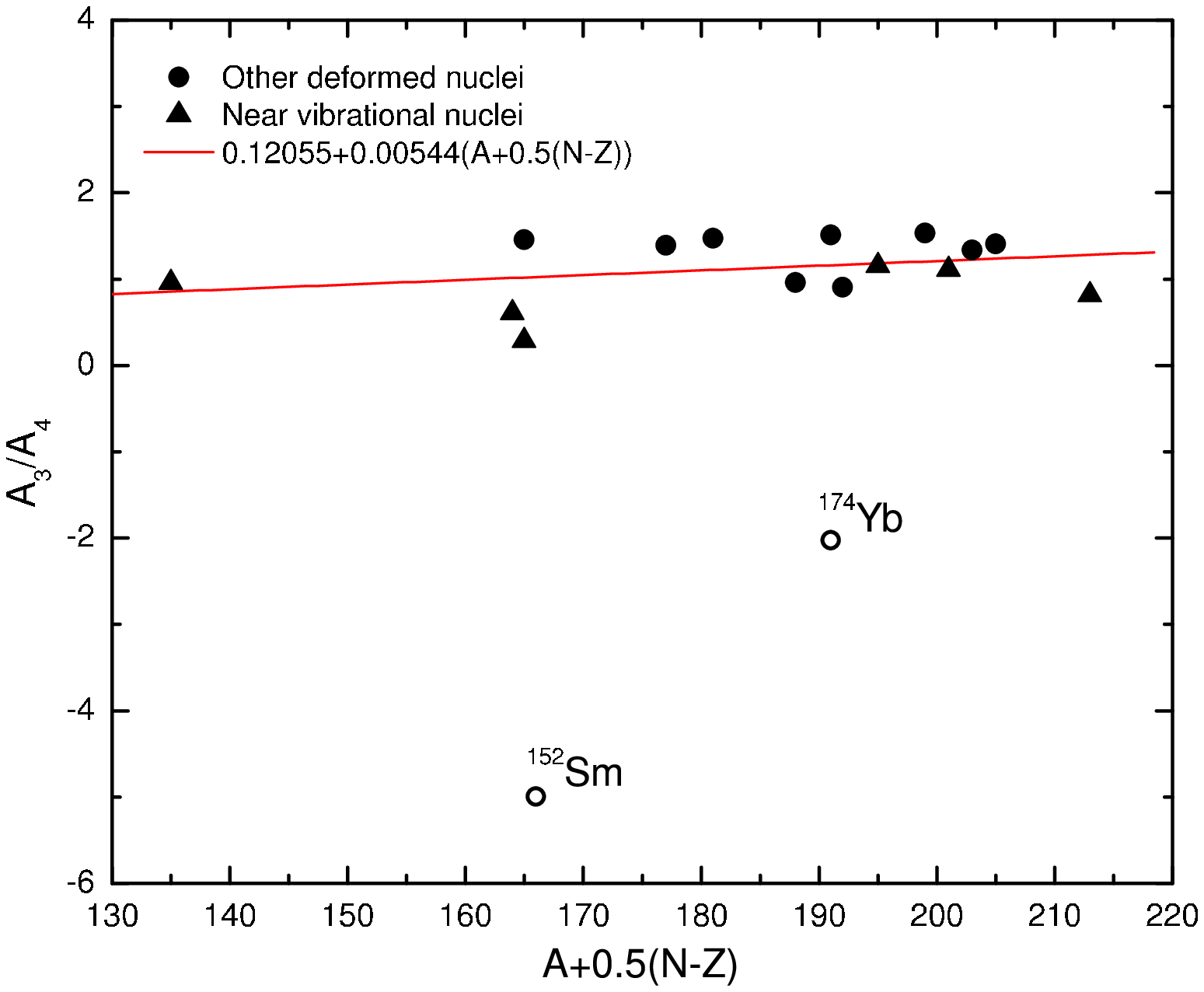}
\end{center}
\caption{For some nuclei a good description of the excitation energies in the beta band is not possible unless the terms $A_3$ and $A_4$ are simultaneously considered. Here we plotted the ratio $A_3/A_4$ as function of $A+(N-Z)/2$.. The resulting points were interpolated by a straight line. }
\label{Fig. 9}
\end{figure}

Now we would like to comment on the change of parameters  when we pass from one nucleus to another. This is done by plotting the deformation parameter as function of the nuclear deformation $\beta_2$, Fig. 5,  and the structure coefficients as functions of $A+(N-Z)/2$, in Figs. 6-9. The atomic mass number was corrected by the third component of the isospin in order to infer also the result dependence on Z. In Fig. 5 it is shown that the three groups of nuclei, near vibrational, transuranic and rare earth deformed nuclei are spread around three distinct straight lines. The lines corresponding to the two groups of well deformed nuclei are almost parallel with each other. The line corresponding to the near vibrational nuclei has a smaller slope than the other two lines. The linear dependence of $d$ on the quadrupole nuclear deformation $\beta_2$ has been studied analytically, in a different context, in Ref. 
\cite{Iudice}. The structure coefficient $A_1$ depends on $A+(N-Z)/2$ according to what is shown in Fig. 6. Again the three groups of nuclei satisfy distinct dependence law. While the transuranic and near vibrational nuclei lie on curves described by fourth order polynomials respectively, the deformed rare earth nuclei belong to a fifth order polynomial curve. Four rare earth nuclei fall apart from the interpolating curve. These are $^{164}$Dy, $^{166}$Er, $^{172}$Yb and $^{174}$Yb.
The parameters $A_2$ yielded by the fitting procedure are interpolated by a sixth order polynomial. Note that the transuranic and the most deformed rare earth nuclei are concentrated around the two minima. reflecting a large moment of inertia for large deformation regime. Below the first minimum five near vibrational nuclei are placed. These nuclei have a small nuclear deformation and therefore the parameter $A_1$ is decisive in determining the excitation energies in the three bands. An interesting dependence on $A+(N-Z)/2$ is shown in Fig. 8 for the structure coefficient $A_3$. Indeed in that case the points corresponding to the  deformed rare earth nuclei cannot be interpolated by a single curve but by two parabolas. Three nuclei deviate substantially from the upper parabola. These are $^{164}$Dy, $^{166}$Er and $^{174}$Yb.
We remember that the band beta is decoupled from the ground and gamma bands. The coupling of this band to the upper bands is simulated by three boson terms having the coefficients $A_3, A_4, A_5$.  For the well deformed nuclei only one term is sufficient in order to obtain a good description for excitation energies in the beta band. This is the term multiplied by $A_3$. However, there are some nuclei where two terms responsible for this band excitations are necessary, $A_3$ and $A_4$, and for few nuclei (5) all three are necessary. We considered all nuclei where $A_4$ term was involved and plotted, in Fig. 9,  the ratio $A_3/A_4$. From there it is seen that this ratio is close to unity. In this plot we didn't consider the case of $^{186}$Hg
where this rule is drastically violated due to the abnormal behavior of $A_5$.
The points of this plot are interpolated with a straight line. Large deviation from this line is noticed for $^{152}$Sm and $^{174}$Yb. The first nucleus plays the role of a critical nucleus in a phase transition from $SU(5)$ to $SU(3)$ symmetry exhibiting $X(5)$ properties. The second nucleus 
seems to be the critical nucleus of a phase transition from a gamma stable to a gamma unstable nucleus, as suggested by Fig. 4. Actually, this assertion is confirmed also by the graphs for $A_1$ and $A_3$ where the interpolating curves have a finite discontinuity. A large deviation from the smooth interpolating curve is seen for the parameter $A_3$ characterizing the isotope $^{162}$Dy.
In the graph associated with $A_1$, the discontinuity is met in $^{164}$Dy and not $^{162}$Dy.
This is caused by the fact that the phase transition in this case is a slow process and not a sharp one.
Coming back to Fig. 4 we note that the excited bands intersect each other. This might be a sign that this nucleus, $^{162}$Dy, is critical for a phase transition from  gamma unstable to a gamma stable regime.
A similar situation is found for $^{166}$Er which is also reflected  in the discontinuities of the 
structure coefficients $A_1$ and $A_3$. In contradistinction to our interpretation in Ref.
\cite{Iache03} the isotopes $^{166}$Er and $^{164}$Dy are considered critical points for the transition from triaxial to axially symmetric shapes, exhibiting the so called Y(5) symmetry. In contrast to the results of Ref. \cite{RaFa} where the isotope $^{154}$Gd is considered to be critical in the transition from the SU(5) to the SU(3) symmetry, here Fig. 8 shows a discontinuity in the $A_3$ behavior for $^{152}$Gd. Indeed, this nucleus is symbolized by the third triangle from the left which stays apart from the rest of the Gd even isotopes represented by the second up to sixth half filled circles from the upper parabola. Note, however, that here $^{152}$Gd is described by a different analytical formula than the rest of the isotopes and moreover by using two additional fitting parameters. However comparing the B(E2) values associated to the transition $2^+_g\to 0^+_g$ in the neighboring isotopes $^{152}$Gd and $^{154}$Gd we notice a ratio larger than 2 in the favor of the second nucleus. Therefore it is more plausible to consider $^{154}$Gd as a critical nucleus. 

Concluding, except for a few nuclei which play the role of critical nuclei for some specific phase transitions all structure coefficients show a smooth dependence on $A+(N-Z)/2$. The deformation parameter $d$ is related linearly with the nuclear deformation $\beta_2$. The coefficients of the linear transformation are different for the three groups of nuclei: near vibrational, rare earth and transuranic nuclei.

\subsection{E2 transition probabilities}
For each considered nucleus the two parameters defining the quadrupole transition operator were determined by a least square fit of the experimental available data.
As mentioned before, for the near vibrational limit the matrix elements of the transition operator $Q_{2\mu}$ were expanded as a power series of $d$ from which we kept  the terms non-depending on d and the next leading order terms which are in most cases linear in $d$. All matrix elements needed for describing the experimental situation were analytically expressed. The B(E2) values are obtained by squaring the reduced matrix elements
obtained as explained before. For near vibrational nuclei the results are collected in Tables V, VI, VII and VIII where, for comparison, the corresponding experimental data are listed. We note that the limit $d\to 0$ provides similar results as the linear expansion in $d$ for the transition operator. However, there are transitions which are forbidden in the vibrational limit but are described quantitatively well by the linear expansion of $Q_{2\mu}$. We remark that for the branching ratios given in Table VIII, the results provided by the vibrational limit are in better agreement with the experimental data than those corresponding to a linear expansion of the transition operator. Moreover, some of the theoretical branching ratios are parameter independent. The agreement between the vibrational limit results and experimental data
is especially good for $^{190}$Pt which is considered to satisfy the $O(6)$ symmetry.

\begin{sidewaystable}[htbp!]
\caption{B(E2) transition probabilities for near vibrational nuclei $^{102}$Pd\cite{DeFrenne}, $^{154}$Dy\cite{Reich1} and $^{152}$Gd\cite{Artna1}. Values in square braces were not taken into account for the fitting procedure.}
\begin{center}
\scriptsize{
\begin{tabular}{|c|c|c|c|c|c|c|c|c|c|c|c|}
\hline
\multicolumn{4}{|c|}{$^{102}$Pd}&\multicolumn{4}{c|}{$^{154}$Dy}&\multicolumn{4}{c|}{$^{152}$Gd}\\
\hline
$B(E2)$ [W.u.]&Exp.&Th.&Th.&$B(E2)$ [W.u.]&Exp.&Th.&Th.&$B(E2)$ [W.u.]&Exp.&Th.&Th.\\
$J_{i}^{\pi}\,\rightarrow\,J_{f}^{\pi}$&&$d\rightarrow0$&Series of $d$&$J_{i}^{\pi}\,\rightarrow\,J_{f}^{\pi}$&&$d\rightarrow0$&Series of $d$&$J_{i}^{\pi}\,\rightarrow\,J_{f}^{\pi}$&&$d\rightarrow0$&Series of $d$\\
\hline
$2_{g}^{+}\rightarrow0_{g}^{+}$&32.6&18.27&20.69   &$2_{g}^{+}\rightarrow0_{g}^{+}$&97&37.20&21.02  &$2_{g}^{+}\rightarrow0_{g}^{+}$&73&60.88&76.00\\
$4_{g}^{+}\rightarrow2_{g}^{+}$&50.9&36.53&43.63   &$4_{g}^{+}\rightarrow2_{g}^{+}$&157&74.40&164.89&$4_{g}^{+}\rightarrow2_{g}^{+}$&134&121.76&140.00\\
\cline{1-4}
$2_{\gamma}^{+}\rightarrow2_{g}^{+}$&15&36.53&24.84&$6_{g}^{+}\rightarrow4_{g}^{+}$&199&111.60&194.50&$6_{g}^{+}\rightarrow4_{g}^{+}$&200&182.65&217.23\\
\cline{9-12}
$2_{\gamma}^{+}\rightarrow0_{g}^{+}$&4.2&4.2&20.09&$8_{g}^{+}\rightarrow6_{g}^{+}$&220&148.80&197.30&$2_{\beta}^{+}\rightarrow0_{\beta}^{+}$&(42)&97.41&24.09\\
\cline{9-12}
                                               &&&&$10_{g}^{+}\rightarrow8_{g}^{+}$&180&186.00&193.21& $0_{\beta}^{+}\rightarrow2_{g}^{+}$&180&180&58.61\\
                                               &&&&$12_{g}^{+}\rightarrow10_{g}^{+}$&170&223.20&191.74&$2_{\beta}^{+}\rightarrow0_{g}^{+}$&(0.31)&0&6.50\\
                                               &&&&$14_{g}^{+}\rightarrow12_{g}^{+}$&200&260.40&194.49&$2_{\beta}^{+}\rightarrow2_{g}^{+}$&(16.9)&0&61.05\\
                                               &&&&                                                &&&&$2_{\beta}^{+}\rightarrow4_{g}^{+}$&[36]&81.00&41.16\\
\hline
\multicolumn{2}{|c|}{$d$                              }&0      &1.45898&\multicolumn{2}{c|}{$d$                              }&0&1.51491&\multicolumn{2}{c|}{$d$                              }&0&1.53241\\
\multicolumn{2}{|c|}{$q_{h}\,[(W.u.)^{\frac{1}{2}}]$  }&4.27391&5.78673&\multicolumn{2}{c|}{$q_{h}\,[(W.u.)^{\frac{1}{2}}]$  }&6.09923&94.8482&\multicolumn{2}{c|}{$q_{h}\,[(W.u.)^{\frac{1}{2}}]$  }&7.80268&2.19266\\
\multicolumn{2}{|c|}{$q_{anh}\,[(W.u.)^{\frac{1}{2}}]$}&1.44928&0.52039&\multicolumn{2}{c|}{$q_{anh}\,[(W.u.)^{\frac{1}{2}}]$}&0&48.8486&\multicolumn{2}{c|}{$q_{anh}\,[(W.u.)^{\frac{1}{2}}]$}&5.47727&6.41363\\
\hline
\end{tabular}}
\end{center}
\end{sidewaystable}

\begin{sidewaystable}[htbp!]
\caption{B(E2) transition probabilities for near vibrational nuclei $^{188}$Os\cite{Balraj1}, $^{190}$Os\cite{Balraj2} and $^{192}$Os\cite{Baglin1}. Values in square braces were not taken into account for the fitting procedure.}
\begin{center}
\scriptsize{
\begin{tabular}{|c|c|c|c|c|c|c|c|c|c|c|c|}
\hline
\multicolumn{4}{|c|}{$^{188}$Os}&\multicolumn{4}{c|}{$^{190}$Os}&\multicolumn{4}{c|}{$^{192}$Os}\\
\hline
$B(E2)$ [W.u.]&Exp.&Th.&Th.&$B(E2)$ [W.u.]&Exp.&Th.&Th.&$B(E2)$ [W.u.]&Exp.&Th.&Th.\\
$J_{i}^{\pi}\,\rightarrow\,J_{f}^{\pi}$&&$d\rightarrow0$&Series of $d$&$J_{i}^{\pi}\,\rightarrow\,J_{f}^{\pi}$&&$d\rightarrow0$&Series of $d$&$J_{i}^{\pi}\,\rightarrow\,J_{f}^{\pi}$&&$d\rightarrow0$&Series of $d$\\
\hline
$2_{g}^{+}\rightarrow0_{g}^{+}$&79    &28.72&27.10&$2_{g}^{+}\rightarrow0_{g}^{+}$&71.9&30.10&22.86&$2_{g}^{+}\rightarrow0_{g}^{+}$&$64.4^{+7}_{-8}$&23.80&23.85\\
$4_{g}^{+}\rightarrow2_{g}^{+}$&133   &57.44 &56.46&$4_{g}^{+}\rightarrow2_{g}^{+}$&105&60.20&46.67&$4_{g}^{+}\rightarrow2_{g}^{+}$&75.6&47.60&49.94\\
$6_{g}^{+}\rightarrow4_{g}^{+}$&(138) &86.15&85.79&$6_{g}^{+}\rightarrow4_{g}^{+}$&113&90.30&70.89&$6_{g}^{+}\rightarrow4_{g}^{+}$&$100^{+5}_{-3}$&71.40&75.72\\
$8_{g}^{+}\rightarrow6_{g}^{+}$&[161] &114.87&114.71&$8_{g}^{+}\rightarrow6_{g}^{+}$&137 &120.40&94.97&$8_{g}^{+}\rightarrow6_{g}^{+}$&115&95.20&101.10\\
$10_{g}^{+}\rightarrow8_{g}^{+}$&[188]&143.59&143.57&$10_{g}^{+}\rightarrow8_{g}^{+}$&[120]&150.50&119.15&$10_{g}^{+}\rightarrow8_{g}^{+}$&$105^{+9}_{-25}$&119.00&126.41\\
\hline
$4_{\gamma}^{+}\rightarrow2_{\gamma}^{+}$&47&45.13&40.19&$4_{\gamma}^{+}\rightarrow2_{\gamma}^{+}$&53&47.30&19.25&$4_{\gamma}^{+}\rightarrow2_{\gamma}^{+}$&($45.3^{+14}_{-18}$)&37.40&41.19\\
$6_{\gamma}^{+}\rightarrow4_{\gamma}^{+}$&[70]&78.32&74.56&$6_{\gamma}^{+}\rightarrow4_{\gamma}^{+}$&[65]&82.09&40.05&$6_{\gamma}^{+}\rightarrow4_{\gamma}^{+}$&($52^{+3}_{-6}$)&64.91&74.41\\
\cline{1-4}
$4_{\gamma}^{+}\rightarrow3_{\gamma}^{+}$&[320]&0&198.11&$8_{\gamma}^{+}\rightarrow6_{\gamma}^{+}$&[61]&114.38&60.73&$8_{\gamma}^{+}\rightarrow6_{\gamma}^{+}$&($48^{+7}_{-6}$)&90.44&105.58\\
\cline{1-12}
$2_{\gamma}^{+}\rightarrow0_{g}^{+}$&5.0    &0.79&10.88&$4_{\gamma}^{+}\rightarrow3_{\gamma}^{+}$&65&0&112.28&$2_{\gamma}^{+}\rightarrow0_{g}^{+}$&$5.62^{+21}_{-12}$&1.73&11.56\\
\cline{5-8}
$2_{\gamma}^{+}\rightarrow2_{g}^{+}$&16&57.44&83.29&$2_{\gamma}^{+}\rightarrow0_{g}^{+}$&5.9&1.09&3.81&$2_{\gamma}^{+}\rightarrow2_{g}^{+}$&$46.0^{+26}_{-12}$&47.60&65.22\\
$2_{\gamma}^{+}\rightarrow4_{g}^{+}$&[34]  &0&14.31&$2_{\gamma}^{+}\rightarrow2_{g}^{+}$&33&60.2&95.13&$4_{\gamma}^{+}\rightarrow2_{g}^{+}$&($0.2876^{+25}_{-34}$)&2.75&0.42\\
$4_{\gamma}^{+}\rightarrow2_{g}^{+}$&1.29&1.25&0.20&$4_{\gamma}^{+}\rightarrow2_{g}^{+}$&0.68&1.71&0.07&$4_{\gamma}^{+}\rightarrow4_{g}^{+}$&$30.9^{+37}_{-18}$&34.00&37.45\\
$4_{\gamma}^{+}\rightarrow4_{g}^{+}$&19&41.03&41.52&$4_{\gamma}^{+}\rightarrow4_{g}^{+}$&30&43.00&31.97&$6_{\gamma}^{+}\rightarrow6_{g}^{+}$&[$26.0^{+55}_{-21}$]&30.29&29.79\\
\cline{9-12}
$4_{\gamma}^{+}\rightarrow6_{g}^{+}$&[16]&0&22.06&$6_{g}^{+}\rightarrow4_{\gamma}^{+}$&[6]&0&9.50&$0_{\beta}^{+}\rightarrow2_{g}^{+}$&[0.60]&5.25&0.61\\
\cline{9-12}
$6_{\gamma}^{+}\rightarrow4_{g}^{+}$&[0.21]&1.70&0.01&$6_{\gamma}^{+}\rightarrow6_{g}^{+}$&[31]&38.31&19.10&$0_{\beta}^{+}\rightarrow2_{\gamma}^{+}$&[$30.4^{+30}_{-23}$]&71.40&66.92\\
\cline{1-8}
$0_{\beta}^{+}\rightarrow2_{g}^{+}$&0.95    &2.38&0.88&$0_{\beta}^{+}\rightarrow2_{g}^{+}$&2.2&3.26&1.33&&&&\\
\cline{1-8}
$0_{\beta}^{+}\rightarrow2_{\gamma}^{+}$&(4.3)    &86.15&64.43&$0_{\beta}^{+}\rightarrow2_{\gamma}^{+}$&(23)&90.30&29.56&&&&\\
\hline
\multicolumn{2}{|c|}{$d$                              }&0      &1.62319 &\multicolumn{2}{c|}{$d$                              }&0&1.59990&\multicolumn{2}{c|}{$d$                              }&0&1.61011\\
\multicolumn{2}{|c|}{$q_{h}\,[(W.u.)^{\frac{1}{2}}]$  }&5.35888&5.02388 &\multicolumn{2}{c|}{$q_{h}\,[(W.u.)^{\frac{1}{2}}]$  }&5.4863&3.85134&\multicolumn{2}{c|}{$q_{h}\,[(W.u.)^{\frac{1}{2}}]$  }&4.87841&4.96308\\
\multicolumn{2}{|c|}{$q_{anh}\,[(W.u.)^{\frac{1}{2}}]$}&0.62933&-0.35919&\multicolumn{2}{c|}{$q_{anh}\,[(W.u.)^{\frac{1}{2}}]$}&0.737623&-0.71724&\multicolumn{2}{c|}{$q_{\renewcommand{\theequation}{2.\arabic{equation}}anh}\,[(W.u.)^{\frac{1}{2}}]$}&0.93545&-0.20690\\
\hline
\end{tabular}}
\end{center}
\end{sidewaystable}

\begin{sidewaystable}[htbp!]
\caption{B(E2) transition probabilities for near vibrational nuclei $^{194}$Pt\cite{Browne1}, $^{196}$Pt\cite{Chun} and $^{186}$Hg\cite{Baglin2}. Values in square braces were not taken into account for the fitting procedure.}
\begin{center}
\scriptsize{
\begin{tabular}{|c|c|c|c|c|c|c|c|c|c|c|c|}
\hline
\multicolumn{4}{|c|}{$^{194}$Pt}&\multicolumn{4}{c|}{$^{196}$Pt}&\multicolumn{4}{c|}{$^{186}$Hg}\\
\hline
$B(E2)$ [W.u.]&Exp.&Th.&Th.&$B(E2)$ [W.u.]&Exp.&Th.&Th.&$B(E2)$ [W.u.]&Exp.&Th.&Th.\\
$J_{i}^{\pi}\,\rightarrow\,J_{f}^{\pi}$&&$d\rightarrow0$&Series of $d$&$J_{i}^{\pi}\,\rightarrow\,J_{f}^{\pi}$&&$d\rightarrow0$&Series of $d$&$J_{i}^{\pi}\,\rightarrow\,J_{f}^{\pi}$&&$d\rightarrow0$&Series of $d$\\
\hline
$2_{g}^{+}\rightarrow0_{g}^{+}$&49.3&27.22 &25.56&$2_{g}^{+}\rightarrow0_{g}^{+}$&40.57&16.23&30.47&$2_{g}^{+}\rightarrow0_{g}^{+}$&44&55.44&57.05\\
\cline{9-12}
$4_{g}^{+}\rightarrow2_{g}^{+}$&85  &54.45 &53.10 &$4_{g}^{+}\rightarrow2_{g}^{+}$&59.9 &32.47&61.57&$2_{\beta}^{+}\rightarrow0_{\beta}^{+}$&[400] &88.71&295.27\\
$6_{g}^{+}\rightarrow4_{g}^{+}$&67  &81.67 &79.20 &$6_{g}^{+}\rightarrow4_{g}^{+}$&[73] &48.70&92.57&$4_{\beta}^{+}\rightarrow2_{\beta}^{+}$&[200] &152.47&286.29\\
$8_{g}^{+}\rightarrow6_{g}^{+}$&50  &108.90&105.89&$8_{g}^{+}\rightarrow6_{g}^{+}$&[78] &64.93&124.39&$6_{\beta}^{+}\rightarrow4_{\beta}^{+}$&290 &211.69&261.19\\
\cline{1-8}
$4_{\gamma}^{+}\rightarrow2_{\gamma}^{+}$&22&42.78&52.31&$4_{\gamma}^{+}\rightarrow2_{\gamma}^{+}$&[29]&25.51&19.46&$8_{\beta}^{+}\rightarrow6_{\beta}^{+}$&$\approx$210 &269.30&239.40\\
\cline{1-4}\cline{9-12}
$2_{\gamma}^{+}\rightarrow0_{g}^{+}$&0.28  &40.23&6.85&$6_{\gamma}^{+}\rightarrow4_{\gamma}^{+}$&49&44.27&41.44&$4_{\beta}^{+}\rightarrow2_{g}^{+}$&80 &0&3.38\\
\cline{5-8}
$2_{\gamma}^{+}\rightarrow2_{g}^{+}$&89    &54.45&50.43&$2_{\beta}^{+}\rightarrow0_{\beta}^{+}$&[5]&25.97&39.80&&&&\\
\cline{5-8}
$4_{\gamma}^{+}\rightarrow2_{g}^{+}$&[0.22]&63.22&0.70&$2_{\gamma}^{+}\rightarrow0_{g}^{+}$&0.0158&0.65&0.04&&&&\\
$4_{\gamma}^{+}\rightarrow4_{g}^{+}$&[20]  &38.89&37.74&$4_{\gamma}^{+}\rightarrow2_{g}^{+}$&[0.56]&1.02&2.18&&&&\\
\cline{1-4}
$0_{\beta}^{+}\rightarrow2_{g}^{+}$&134&120.68&1.53&$6_{\gamma}^{+}\rightarrow4_{g}^{+}$&0.48&1.39&4.08&&&&\\
\cline{1-8}
$0_{\beta}^{+}\rightarrow2_{\gamma}^{+}$&135&81.67&89.62&$0_{\beta}^{+}\rightarrow2_{g}^{+}$&2.8     &1.94&13.81&&&&\\
                                                          &&&&$2_{\beta}^{+}\rightarrow0_{g}^{+}$&[0.0025]&0     &0.05&&&&\\
                                                          &&&&$2_{\beta}^{+}\rightarrow4_{g}^{+}$&[0.13]  &2.43&6.39&&&&\\
\cline{5-8}
                                                          &&&&$0_{\beta}^{+}\rightarrow2_{\gamma}^{+}$&18&48.70&29.80&&&&\\
\hline
\multicolumn{2}{|c|}{$d$                              }&0      &0.83137&\multicolumn{2}{c|}{$d$                              }&0&0.95083&\multicolumn{2}{c|}{$d$                              }&0&0.92388\\
\multicolumn{2}{|c|}{$q_{h}\,[(W.u.)^{\frac{1}{2}}]$  }&5.21769&5.46832&\multicolumn{2}{c|}{$q_{h}\,[(W.u.)^{\frac{1}{2}}]$  }&4.02904&4.13681&\multicolumn{2}{c|}{$q_{h}\,[(W.u.)^{\frac{1}{2}}]$  }&7.44606&34.2497\\
\multicolumn{2}{|c|}{$q_{anh}\,[(W.u.)^{\frac{1}{2}}]$}&4.48486&0.09065&\multicolumn{2}{c|}{$q_{anh}\,[(W.u.)^{\frac{1}{2}}]$}&0.56888&-1.43127&\multicolumn{2}{c|}{$q_{anh}\,[(W.u.)^{\frac{1}{2}}]$}&0&35.7881\\
\hline
\end{tabular}}
\end{center}
\end{sidewaystable}

\begin{sidewaystable}[htbp!]
\caption{Branching ratios for near vibrational nuclei $^{126}$Xe\cite{Katakura}, $^{182}$Pt\cite{Balraj3}, $^{186}$Pt\cite{Baglin2} and $^{190}$Pt \cite{Finger}.}
\begin{center}
\scriptsize{
\begin{tabular}{|c|c|c|c|c|c|c|c|c|c|c|c|c|c|c|c|}
\hline
\multicolumn{4}{|c|}{$^{126}$Xe }&\multicolumn{4}{c|}{$^{182}$Pt}&\multicolumn{4}{c|}{$^{186}$Pt}&\multicolumn{4}{c|}{$^{190}$Pt}\\
\hline
$\frac{B(E2;I\rightarrow I')}{B(E2;J\rightarrow J')}$ &Exp.&Th.&Th.&$\frac{B(E2;I\rightarrow I')}{B(E2;J\rightarrow J')}$&Exp.&Th.&Th.&$\frac{B(E2;I\rightarrow I')}{B(E2;J\rightarrow J')}$&Exp.&Th.&Th.&$\frac{B(E2;I\rightarrow I')}{B(E2;J\rightarrow J')}$&Exp.&Th.&Th\\
$\times10^{-2}$&&$d\rightarrow0$&Series of $d$&&&$d\rightarrow0$&Series of $d$&$\times10^{-2}$&&$d\rightarrow0$&Series of $d$&$\times10^{-2}$&&$d\rightarrow0$&Series of $d$\\
\hline
$\frac{2_{\gamma}^{+}\rightarrow0_{g}^{+}}{2_{\gamma}^{+}\rightarrow2_{g}^{+}}$          &1.38 &1.63        &2.96 &$\frac{2_{\beta}^{+}\rightarrow2_{g}^{+}}{2_{\beta}^{+}\rightarrow0_{g}^{+}}$  &0.24   &0   &11.52&$\frac{2_{\gamma}^{+}\rightarrow0_{g}^{+}}{2_{\gamma}^{+}\rightarrow2_{g}^{+}}$     &9.31 &9.31       &34.05     &$\frac{2_{\gamma}^{+}\rightarrow0_{g}^{+}}{2_{\gamma}^{+}\rightarrow2_{g}^{+}}$     &1.24&2.86    &4.80\\
$\frac{0_{\beta}^{+}\rightarrow2_{g}^{+}}{0_{\beta}^{+}\rightarrow2_{\gamma}^{+}}$       &9.27 &3.25        &9.25 &$\frac{2_{\gamma}^{+}\rightarrow2_{g}^{+}}{2_{\gamma}^{+}\rightarrow0_{g}^{+}}$&$<$4.94&4.94&12.67&$\frac{2_{\beta}^{+}\rightarrow0_{g}^{+}}{2_{\beta}^{+}\rightarrow0_{\beta}^{+}}$   &7.59 &0          &$\approx0$&$\frac{3_{\gamma}^{+}\rightarrow4_{g}^{+}}{3_{\gamma}^{+}\rightarrow2_{\gamma}^{+}}$&49  &40$^{*}$&20.28\\
$\frac{3_{\gamma}^{+}\rightarrow4_{g}^{+}}{3_{\gamma}^{+}\rightarrow2_{\gamma}^{+}}$     &85.60&40$^{*}$    &20.34&                                                                               &       &    &     &$\frac{2_{\beta}^{+}\rightarrow0_{g}^{+}}{2_{\beta}^{+}\rightarrow4_{g}^{+}}$       &11.06&0          &0.03      &$\frac{3_{\gamma}^{+}\rightarrow2_{g}^{+}}{3_{\gamma}^{+}\rightarrow2_{\gamma}^{+}}$&1.8 &5.73    &11.83\\
$\frac{3_{\gamma}^{+}\rightarrow2_{g}^{+}}{3_{\gamma}^{+}\rightarrow2_{\gamma}^{+}}$     &2.29 &3.25        &7.50 &                                                                               &       &    &     &$\frac{4_{\gamma}^{+}\rightarrow4_{g}^{+}}{4_{\gamma}^{+}\rightarrow2_{\gamma}^{+}}$&42.03&90.91$^{*}$&73.44     &$\frac{0_{\beta}^{+}\rightarrow2_{g}^{+}}{0_{\beta}^{+}\rightarrow2_{\gamma}^{+}}$  &11  &16.20   &2.63\\
$\frac{4_{\gamma}^{+}\rightarrow4_{g}^{+}}{4_{\gamma}^{+}\rightarrow2_{\gamma}^{+}}$     &86.33&90.91$^{*}$ &93.98&                                                                               &       &    &     &$\frac{4_{\beta}^{+}\rightarrow2_{g}^{+}}{4_{\beta}^{+}\rightarrow2_{\beta}^{+}}$   &2.61 &0          &$\approx0$&$\frac{2_{\beta}^{+}\rightarrow0_{g}^{+}}{2_{\beta}^{+}\rightarrow0_{\beta}^{+}}$   &0.2 &0       &$\approx0$\\
$\frac{4_{\gamma}^{+}\rightarrow2_{g}^{+}}{4_{\gamma}^{+}\rightarrow2_{\gamma}^{+}}$     &1.09 &3.25        &0.18 &                                                                               &       &    &     &                                                                                    &     &           &          &$\frac{2_{\beta}^{+}\rightarrow4_{g}^{+}}{2_{\beta}^{+}\rightarrow0_{\beta}^{+}}$   &4.2 &4.83    &4.13\\
$\frac{5_{\gamma}^{+}\rightarrow4_{\gamma}^{+}}{5_{\gamma}^{+}\rightarrow3_{\gamma}^{+}}$&94.04&45.45$^{*}$ &46.20&                                                                               &       &    &     &                                                                                    &     &           &&&&&\\
$\frac{5_{\gamma}^{+}\rightarrow4_{g}^{+}}{5_{\gamma}^{+}\rightarrow3_{\gamma}^{+}}$     &3.76 &4.64        &1.79 &&&&&&&&&&&&\\
$\frac{6_{\gamma}^{+}\rightarrow6_{g}^{+}}{6_{\gamma}^{+}\rightarrow4_{\gamma}^{+}}$     &86.19&46.67$^{*}$ &41.16&&&&&&&&&&&&\\
$\frac{6_{\gamma}^{+}\rightarrow4_{g}^{+}}{6_{\gamma}^{+}\rightarrow4_{\gamma}^{+}}$     &0.63 &2.55        &0.45 &&&&&&&&&&&&\\
\hline
\multicolumn{2}{|c|}{$d$             }&0&0.67610 &\multicolumn{2}{c|}{$d$             }&0&1.69634&\multicolumn{2}{c|}{$d$             }&0&1.67744&\multicolumn{2}{c|}{$d$             }&0&0.74815\\
\multicolumn{2}{|c|}{$q_{anh}/q_{h}$}&0.12749&-0.10137&\multicolumn{2}{c|}{$q_{anh}/q_{h}$}&0.450&0.60093&\multicolumn{2}{c|}{$q_{anh}/q_{h}$}&0.30512&0.00418&\multicolumn{2}{c|}{$q_{anh}/q_{h}$}&0.16922&-0.07169\\
\hline
\end{tabular}}
\end{center}

$^{*}$ Do not depend on any parameters.
\end{sidewaystable}

The B(E2) values for the well deformed nuclei considered by the present work, have been calculated using the asymptotic expressions for the matrix elements given by Eqs.(\ref{Alag1}) and 
(\ref{Alag2}). The results are listed in Tables IX, X, XI and XII. As seen from these tables, a very good agreement between the results of our calculations and the corresponding experimental data is obtained. A special mention is deserved by  $^{156}$Gd, $^{158}$Gd, $^{152}$Sm and $^{232}$Th where 25, 23, 22 and 20 B(E2) values  are available  respectively, and an excellent agreement is obtained. Also for $^{172}$Yb, $^{182}$W and $^{186}$Os, 17, 18 and 17 transitions respectively, are known and all of them are nicely described by the formalism proposed.

It is worth mentioning that the list of nuclei considered in the present paper include isotopes of various "nuclear phases" with specific symmetries like 
gamma stable, gamma unstable, triaxial shape, deformed axial symmetric nuclei showing a SU(3) symmetry, critical nuclei of various phase transitions 
satisfying the symmetries $E(5)$ ($^{102}$Pd) and $X(5)$ ($^{152}$Sm, $^{154}$Gd) respectively.  
In the isotopic chain of $^{152-162}$Gd, two phase transitions take place namely from SU(5) to SU(3)
with the critical nucleus $^{154}$Gd and from SU(3) to O(6), i.e. to a gamma unstable shape,  the critical nucleus being $^{160}$Gd \cite{RaFa}.
The properties of all these nuclei can be described fairly well by CSM. Moreover, in this work we present analytical expressions for energies and transition probabilities for vibrational, transitional and well deformed nuclei. Since the analytical formulas are easy to be handled, and by this paper they were positively tested, we may say that the results presented here represent a
major achievement of CSM.

\begin{table}[htbp!]
\caption{B(E2) transition probabilities in the asymptotic limit for deformed Gd nuclei. Values in square braces were not taken into account for the fitting procedure. Experimental data are taken from \cite{Artna1,Reich1,Reich2,Helmer1,Reich3,Helmer2}.}
\begin{center}
\scriptsize{
\begin{tabular}{|c|c|c|c|c|c|c|c|c|}
\hline
B(E2) tr. prob.&\multicolumn{2}{c|}{$^{154}$Gd}&\multicolumn{2}{c|}{$^{156}$Gd}&\multicolumn{2}{c|}{$^{158}$Gd}&\multicolumn{2}{c|}{$^{160}$Gd}\\
\hline
$J_{i}^{\pi}\,\rightarrow\,J_{f}^{\pi}$&Exp.&Th.&Exp.&Th.&Exp.&Th.&Exp.&Th.\\
\hline
$2_{g}^{+}\rightarrow0_{g}^{+}$  &157  &161.4021&187&181.9295&198  &200.4643&201.2&200.5707\\
$4_{g}^{+}\rightarrow2_{g}^{+}$  &245  &230.5744&263&259.8993&289  &286.3776&     &        \\
$6_{g}^{+}\rightarrow4_{g}^{+}$  &285  &253.9543&295&286.2527&     &        &     &        \\
$8_{g}^{+}\rightarrow6_{g}^{+}$  &[312]&265.8387&320&299.6486&[330]&330.1765&     &        \\
$10_{g}^{+}\rightarrow8_{g}^{+}$ &     &        &314&307.7755&340  &339.1314&     &        \\
$12_{g}^{+}\rightarrow10_{g}^{+}$&     &        &300&313.2351&[310]&345.1473&     &        \\
\hline
$2_{\beta}^{+}\rightarrow0_{\beta}^{+}$&97&161.4021&[52]&181.9295&     &        &&\\
$4_{\beta}^{+}\rightarrow2_{\beta}^{+}$&  &        &280 &259.8993&[455]&286.3776&&\\
\hline
$4_{\gamma}^{+}\rightarrow2_{\gamma}^{+}$&&&               &        &[113]&119.3240&&\\
$5_{\gamma}^{+}\rightarrow3_{\gamma}^{+}$&&&100$^{+3}_{-1}$&173.6600&     &        &&\\
\hline
$0_{\beta}^{+}\rightarrow2_{g}^{+}$&52  &46.2919&8   &7.4675&1.1652&2.2242&&\\
$2_{\beta}^{+}\rightarrow0_{g}^{+}$&0.86&9.2584 &0.63&1.4935&0.31  &0.4448&&\\
$2_{\beta}^{+}\rightarrow2_{g}^{+}$&6.7 &13.2263&3.3 &2.1336&0.079 &0.6355&&\\
$2_{\beta}^{+}\rightarrow4_{g}^{+}$&19.6&23.8073&4.1 &3.8404&1.39  &1.1439&&\\
$4_{\beta}^{+}\rightarrow2_{g}^{+}$&    &       &1.3 &2.1336&1.32  &0.6355&&\\
$4_{\beta}^{+}\rightarrow4_{g}^{+}$&    &       &    &      &0.37  &0.5777&&\\
$4_{\beta}^{+}\rightarrow6_{g}^{+}$&    &       &2.1 &3.3943&3.16  &1.0110&&\\
\hline
$2_{\gamma}^{+}\rightarrow0_{g}^{+}$&5.7 &10.8613&4.68            &9.5440 &3.4    &9.1615 &3.80&9.1323\\
$2_{\gamma}^{+}\rightarrow2_{g}^{+}$&12.3&15.5162&7.24            &13.6343&6.0    &13.0879&7.1 &13.0461\\
$2_{\gamma}^{+}\rightarrow4_{g}^{+}$&1.72&0.7758 &0.77            &0.6817 &(0.27) &0.6544 &0.72&0.6523\\
$3_{\gamma}^{+}\rightarrow2_{g}^{+}$&    &       &7.3             &17.0428&3.5    &16.3599&    & \\
$3_{\gamma}^{+}\rightarrow4_{g}^{+}$&    &       &5.1             &6.8171 &1.77   &6.5439 &    & \\
$4_{\gamma}^{+}\rightarrow2_{g}^{+}$&    &       &1.8             &5.6809 &1.13   &5.4533 &    &\\
$4_{\gamma}^{+}\rightarrow4_{g}^{+}$&    &       &10              &16.7329&7.31   &16.0624&    &\\
$4_{\gamma}^{+}\rightarrow6_{g}^{+}$&    &       &                &       &[0.949]& 1.3881&    & \\
$5_{\gamma}^{+}\rightarrow4_{g}^{+}$&    &       &8$^{+16}_{-8}$  &15.1836&       &       &    &\\
$5_{\gamma}^{+}\rightarrow6_{g}^{+}$&    &       &11$^{+23}_{-11}$&8.6763 &       &       &    &\\
\hline
$2_{\gamma}^{+}\rightarrow0_{\beta}^{+}$&[1.21]&0.1270&&&&&&\\
$4_{\gamma}^{+}\rightarrow2_{\beta}^{+}$&&&[4.3]&0.0664&&&&\\
$4_{\beta}^{+}\rightarrow2_{\gamma}^{+}$&&&&&[12.8]&0.0043&&\\
\hline
$d$&\multicolumn{2}{c|}{2.72583}&\multicolumn{2}{c|}{3.08725}&\multicolumn{2}{c|}{3.30765}&\multicolumn{2}{c|}{3.31382}\\
$q_{h}\,[(W.u.)^{\frac{1}{2}}]$&\multicolumn{2}{c|}{5.21088}&\multicolumn{2}{c|}{4.88466}&\multicolumn{2}{c|}{4.78579}&\multicolumn{2}{c|}{4.77815}\\
$q_{anh}\,[(W.u.)^{\frac{1}{2}}]$&\multicolumn{2}{c|}{5.60468}&\multicolumn{2}{c|}{2.25107}&\multicolumn{2}{c|}{1.22853}&\multicolumn{2}{c|}{0}\\
\hline
\end{tabular}}
\end{center}
\end{table}

\begin{table}[htbp!]
\caption{B(E2) transition probabilities in the asymptotic limit for few deformed transuranic nuclei which have absolute experimental values also for inter-band transitions. Values in square braces were not taken into account for the fitting procedure. Only for $^{230}$Th and $^{238}$Pu were used the uncertain $\beta$-ground transition probabilities in order to fix the $q_{anh}$ parameter. Experimental data are taken from \cite{Akovali1,Schmorak1,Chukreev1}.}
\begin{center}
\scriptsize{
\begin{tabular}{|c|c|c|c|c|c|c|c|c|c|}
\hline
B(E2) tr. prob.&\multicolumn{2}{c|}{$^{230}$Th}&\multicolumn{2}{c|}{$^{232}$Th}&\multicolumn{2}{c|}{$^{238}$U}&\multicolumn{2}{c|}{$^{238}$Pu}\\
\hline
$J_{i}^{\pi}\,\rightarrow\,J_{f}^{\pi}$&Exp.&Th.&Exp.&Th.&Exp.&Th.&Exp.&Th.\\
\hline
$2_{g}^{+}\rightarrow0_{g}^{+}$     &192   &185.3438&198          &223.2176&281                &280.2843&285  &285\\
$4_{g}^{+}\rightarrow2_{g}^{+}$     &261   &264.7768&286          &318.8822&                   &        &     &\\
$6_{g}^{+}\rightarrow4_{g}^{+}$     &      &        &327          &351.2164&                   &        &     &\\
$8_{g}^{+}\rightarrow6_{g}^{+}$     &      &        &343          &367.6524&$[404^{+67}_{-47}]$&400.4061&     &\\
$10_{g}^{+}\rightarrow8_{g}^{+}$    &      &        &361          &377.6237&$[480^{+61}_{-48}]$&474.1651&     &\\
$12_{g}^{+}\rightarrow10_{g}^{+}$   &      &        &370          &384.3224&[500]              &482.5765&     &\\
$14_{g}^{+}\rightarrow12_{g}^{+}$   &      &        &390          &389.1341&[491]              &488.6182&     &\\
$16_{g}^{+}\rightarrow14_{g}^{+}$   &      &        &390          &392.7582&                   &        &     &\\
$18_{g}^{+}\rightarrow16_{g}^{+}$   &      &        &440          &395.5863&[480]              &496.7200&     &\\
$20_{g}^{+}\rightarrow18_{g}^{+}$   &      &        &360          &397.8550&[460]              &499.5687&     &\\
$22_{g}^{+}\rightarrow20_{g}^{+}$   &      &        &420          &399.7152&[490]              &501.9044&     &\\
$24_{g}^{+}\rightarrow22_{g}^{+}$   &      &        &240          &401.2682&                   &        &     &\\
$26_{g}^{+}\rightarrow24_{g}^{+}$   &      &        &350          &402.5844&                   &        &     &\\
$28_{g}^{+}\rightarrow26_{g}^{+}$   &      &        &705          &403.7141&                   &        &     &\\
\hline
$2_{\beta}^{+}\rightarrow0_{g}^{+}$ &[1.1] &1.1223  &2.3          &1.0374  &[0.38]             &0.7     &[3.9]&1.5595\\
$2_{\beta}^{+}\rightarrow2_{g}^{+}$ &      &        &$\approx0.$  &1.4820  &1.0                &1.0     &     &\\
$2_{\beta}^{+}\rightarrow4_{g}^{+}$ &[3.8] &2.8859  &[$\approx3.$]&2.6676  &[3.3]              &1.8     &[3.1]&4.0102\\
\hline
$2_{\gamma}^{+}\rightarrow0_{g}^{+}$&3.0   &8.9506  &3.0          &9.8088  &3.04               &10.0168 &     &\\
$2_{\gamma}^{+}\rightarrow2_{g}^{+}$&5.4   &12.7866 &7.1          &14.0126 &5.3                &14.3097 &     &\\
$2_{\gamma}^{+}\rightarrow4_{g}^{+}$&[0.35]&0.6393  &$\approx0.$  &0.7006  &0.33               &0.7155  &     &\\
\hline
$d$                              &\multicolumn{2}{c|}{3.21904}&\multicolumn{2}{c|}{3.37319}&\multicolumn{2}{c|}{3.74042}&\multicolumn{2}{c|}{3.96825}\\
$q_{h}\,[(W.u.)^{\frac{1}{2}}]$  &\multicolumn{2}{c|}{4.73039}&\multicolumn{2}{c|}{4.95197}&\multicolumn{2}{c|}{5.00419}&\multicolumn{2}{c|}{4.75640}\\
$q_{anh}\,[(W.u.)^{\frac{1}{2}}]$&\multicolumn{2}{c|}{1.95136}&\multicolumn{2}{c|}{2.25107}&\multicolumn{2}{c|}{1.54111}&\multicolumn{2}{c|}{2.30027}\\
\hline
\end{tabular}}
\end{center}
\end{table}

\begin{table}[htbp!]
\caption{B(E2) transition probabilities in the asymptotic limit for lightest rare earth nuclei. Values in square braces were not taken into account for the fitting procedure. The uncertain transition probabilities were used for $^{150}$Nd in order made the fit. Experimental data are taken from \cite{derMat,Artna1,Helmer3,Balraj4,Shursh}.}
\begin{center}
\scriptsize{
\begin{tabular}{|c|c|c|c|c|c|c|c|c|c|c|}
\hline
B(E2) tr. prob.&\multicolumn{2}{c|}{$^{150}$Nd}&\multicolumn{2}{c|}{$^{152}$Sm}&\multicolumn{2}{c|}{$^{162}$Dy}&\multicolumn{2}{c|}{$^{164}$Dy}&\multicolumn{2}{c|}{$^{166}$Er}\\
\hline
$J_{i}^{\pi}\,\rightarrow\,J_{f}^{\pi}$&Exp.&Th.&Exp.&Th.&Exp.&Th.&Exp.&Th.&Exp.&Th.\\
\hline
$2_{g}^{+}\rightarrow0_{g}^{+}$          &[115]   &131.9577&144           &183.2375&199       &192.5001&209 &198.8182&214 &219.0829\\
$4_{g}^{+}\rightarrow2_{g}^{+}$          &[175]   &188.5110&209           &261.7679&288       &275.0002&272 &284.0207&311 &312.9756\\
$6_{g}^{+}\rightarrow4_{g}^{+}$          &[212]   &207.6258&245           &288.3108&300       &302.8848&325 &312.8259&347 &344.9756\\
$8_{g}^{+}\rightarrow6_{g}^{+}$          &[236]   &217.3421&285           &301.8030&[347]     &317.0590&310 &327.4653&365 &360.8425\\
$10_{g}^{+}\rightarrow8_{g}^{+}$         &        &        &320           &309.9883&[350]     &325.6581&354 &336.3466&371 &370.6290\\
$12_{g}^{+}\rightarrow10_{g}^{+}$        &        &        &              &        &320       &331.4350&356 &342.3132&376 &377.2037\\
$14_{g}^{+}\rightarrow12_{g}^{+}$        &        &        &              &        &330       &335.5845&326 &346.5988&    &        \\
\hline
$2_{\beta}^{+}\rightarrow0_{\beta}^{+}$  &        &        &[520]         &183.2375&          &        &    &        &    &        \\
$4_{\beta}^{+}\rightarrow2_{\beta}^{+}$  &        &        &$\approx400$  &261.7679&          &        &    &        &    &        \\
\hline
$4_{\gamma}^{+}\rightarrow2_{\gamma}^{+}$&        &        &[50]          &109.0700&          &        &    &        &    &        \\
\hline
$0_{\beta}^{+}\rightarrow2_{g}^{+}$      &[0.0428]&8.9884  &32.7          &28.2136 &          &        &    &        &    &        \\
\hline
$2_{\beta}^{+}\rightarrow0_{g}^{+}$      &[0.51]  &1.7977  &0.92          &5.6427  &          &        &    &        &    &        \\
$2_{\beta}^{+}\rightarrow2_{g}^{+}$      &[7.1]   &2.5681  &5.5           &8.0610  &          &        &    &        &    &        \\
$2_{\beta}^{+}\rightarrow4_{g}^{+}$      &[20]    &4.6226  &(19.0)        &14.5098 &          &        &    &        &    &        \\
$4_{\beta}^{+}\rightarrow2_{g}^{+}$      &        &        &$\approx1.0$  &8.0610  &          &        &    &        &    &        \\
$4_{\beta}^{+}\rightarrow4_{g}^{+}$      &        &        &$\approx9.0$  &7.3282  &          &        &    &        &    &        \\
$4_{\beta}^{+}\rightarrow6_{g}^{+}$      &        &        &$(\approx6.0)$&12.8243 &          &        &    &        &    &        \\
\hline
$2_{\gamma}^{+}\rightarrow0_{g}^{+}$     &[3.0]   &9.7248  &3.62          &12.8838 &[0.0241]  &8.9188  &4.0 &10.6601 &5.5 &12.6207 \\
$2_{\gamma}^{+}\rightarrow2_{g}^{+}$     &[5.7]   &13.8926 &9.3           &18.4054 &$\approx0$&12.7411 &8.0 &15.2288 &9.7 &18.0296 \\
$2_{\gamma}^{+}\rightarrow4_{g}^{+}$     &[1.7]   &0.6946  &(0.78)        &0.9203  &[0.00330] &0.6371  &0.96&0.7614  &0.67&0.9015  \\
$4_{\gamma}^{+}\rightarrow2_{g}^{+}$     &        &        &0.59          &7.6689  &          &        &    &        &    &        \\
$4_{\gamma}^{+}\rightarrow4_{g}^{+}$     &        &        &5.5           &22.5885 &          &        &    &        &    &        \\
$4_{\gamma}^{+}\rightarrow6_{g}^{+}$     &        &        &[1.2]         &1.9521  &          &        &    &        &    &        \\
\hline
$4_{\gamma}^{+}\rightarrow2_{\beta}^{+}$ &        &        &[0.18]        &0.0897  &          &        &    &        &    &        \\
\hline
$d$                              &\multicolumn{2}{c|}{2.60473}&\multicolumn{2}{c|}{2.66667}&\multicolumn{2}{c|}{3.28509}&\multicolumn{2}{c|}{3.05374}&\multicolumn{2}{c|}{2.94610}\\
$q_{h}\,[(W.u.)^{\frac{1}{2}}]$  &\multicolumn{2}{c|}{4.93072}&\multicolumn{2}{c|}{5.67534}&\multicolumn{2}{c|}{4.72197}&\multicolumn{2}{c|}{5.16240}&\multicolumn{2}{c|}{5.61710}\\
$q_{anh}\,[(W.u.)^{\frac{1}{2}}]$&\multicolumn{2}{c|}{2.46967}&\multicolumn{2}{c|}{4.37549}&\multicolumn{2}{c|}{0}      &\multicolumn{2}{c|}{0}      &\multicolumn{2}{c|}{0}\\
\hline
\end{tabular}}
\end{center}
\end{table}

\begin{table}[htbp!]
\caption{B(E2) transition probabilities in the asymptotic limit for heaviest rare earth nuclei. Values in square braces were not taken into account for the fitting procedure. Experimental data are taken from \cite{Balraj3,Baglin2,Balraj5,Browne2,Browne3,Browne4}.}
\begin{center}
\scriptsize{
\begin{tabular}{|c|c|c|c|c|c|c|c|c|c|c|c|c|}
\hline
B(E2) tr. prob.&\multicolumn{2}{c|}{$^{172}$Yb}&\multicolumn{2}{c|}{$^{174}$Yb}&\multicolumn{2}{c|}{$^{176}$Hf}&\multicolumn{2}{c|}{$^{182}$W}&\multicolumn{2}{c|}{$^{182}$W}&\multicolumn{2}{c|}{$^{186}$Os}\\
\hline
$J_{i}^{\pi}\,\rightarrow\,J_{f}^{\pi}$&Exp.&Th.&Exp.&Th.&Exp.&Th.&Exp.&Th.&Exp.&Th.&Exp.&Th.\\
\hline
$2_{g}^{+}\rightarrow0_{g}^{+}$           &212                           &226.1302&201               &214.4049&183&182.7660&137   &119.5874&111              &100.9240&92.3 &94.6028\\
$4_{g}^{+}\rightarrow2_{g}^{+}$           &301                           &323.0431&280               &306.2927&   &        &196   &170.8392&144              &144.1772&134  &135.1469\\
$6_{g}^{+}\rightarrow4_{g}^{+}$           &320                           &355.7992&370               &337.3503&   &        &200   &188.1620&187              &158.7966&184  &148.8506\\
$8_{g}^{+}\rightarrow6_{g}^{+}$           &400                           &372.4497&[388]             &353.1374&   &        &209   &196.9675&178              &166.2278&174  &155.8164\\
$10_{g}^{+}\rightarrow8_{g}^{+}$          &375                           &382.5510&[335]             &362.7150&   &        &203   &202.3095&$151^{+15}_{-45}$&170.7362&190  &160.0424\\
$12_{g}^{+}\rightarrow10_{g}^{+}$         &(400)                         &389.3372&[369]             &369.1493&   &        &191   &205.8983&$189^{+20}_{-56}$&173.7649&170  &162.8814\\
$14_{g}^{+}\rightarrow12_{g}^{+}$         &$\left(394^{+60}_{-45}\right)$&394.2116&[320]             &373.7709&   &        &(170) &208.4761&138              &175.9404&     &\\
$16_{g}^{+}\rightarrow14_{g}^{+}$         &                              &        &                  &        &   &        &[204] &210.4177&                 &        &     &\\
$18_{g}^{+}\rightarrow16_{g}^{+}$         &                              &        &                  &        &   &        &[250] &211.9329&                 &        &     &\\
\hline
$2_{\beta}^{+}\rightarrow0_{\beta}^{+}$   &                              &        &                  &        &   &        &[200] &119.5874&                 &        &     &\\
\hline
$4_{\gamma}^{+}\rightarrow2_{\gamma}^{+}$ &                              &        &                  &        &   &        &      &        &                 &        &72   &56.3112\\
$6_{\gamma}^{+}\rightarrow4_{\gamma}^{+}$ &                              &        &                  &        &   &        &      &        &                 &        &119  &111.1418\\
$8_{\gamma}^{+}\rightarrow6_{\gamma}^{+}$ &                              &        &                  &        &   &        &      &        &                 &        &79   &134.1532\\
$10_{\gamma}^{+}\rightarrow8_{\gamma}^{+}$&                              &        &                  &        &   &        &      &        &                 &        &89   &146.0535\\
\hline
$0_{\beta}^{+}\rightarrow2_{g}^{+}$       &[3.6]                         &4.0387  &$[1.4^{+11}_{-5}]$&1.4     &   &        &      &        &                 &        &     &\\
\hline
$2_{\beta}^{+}\rightarrow0_{g}^{+}$       &0.24                          &0.8077  &                  &        &1.0&2.0568  &0.91  &0.6483  &                 &        &     &\\
$2_{\beta}^{+}\rightarrow2_{g}^{+}$       &0.79                          &1.1539  &                  &        &   &        &0.63  &0.9262  &                 &        &     &\\
$2_{\beta}^{+}\rightarrow4_{g}^{+}$       &2.5                           &2.0770  &                  &        &5.7&5.2890  &1.73  &1.6672  &                 &        &     &\\
\hline
$2_{\gamma}^{+}\rightarrow0_{g}^{+}$      &1.33                          &8.2744  &                  &        &4.1&8.9042  &3.40  &5.8162  &4.63             &7.5447  &10.1 &8.3604\\
$2_{\gamma}^{+}\rightarrow2_{g}^{+}$      &                              &        &2.5               &10.6707 &   &        &6.74  &8.3089  &10.1             &10.7781 &23.5 &11.9435\\
$2_{\gamma}^{+}\rightarrow4_{g}^{+}$      &(0.129)                       &0.5910  &                  &        &   &        &0.0339&0.4154  &                 &        &[1.2]&0.5972\\
$4_{\gamma}^{+}\rightarrow2_{g}^{+}$      &7                             &4.9252  &                  &        &   &        &2.35  &3.4620  &                 &        &3.2  &4.9764\\
$4_{\gamma}^{+}\rightarrow4_{g}^{+}$      &13                            &14.5070 &                  &        &   &        &10.4  &10.1973 &                 &        &24.7 &14.6579\\
$6_{\gamma}^{+}\rightarrow4_{g}^{+}$      &                              &        &                  &        &   &        &      &        &                 &        &1.27 &4.0925\\
$6_{\gamma}^{+}\rightarrow6_{g}^{+}$      &                              &        &                  &        &   &        &      &        &                 &        &18.5 &15.2008\\
\hline
$2_{\gamma}^{+}\rightarrow0_{\beta}^{+}$  &[2.42]                        &1.3344  &                  &        &   &        &      &        &                 &        &     &\\
$2_{\gamma}^{+}\rightarrow2_{\beta}^{+}$  &[3.4]                         &1.9064  &                  &        &   &        &      &        &                 &        &     &\\
\hline
$d$                              &\multicolumn{2}{c|}{3.69655}&\multicolumn{2}{c|}{3.78841}&\multicolumn{2}{c|}{3.20357}&\multicolumn{2}{c|}{3.20632}&\multicolumn{2}{c|}{2.58620}&\multicolumn{2}{c|}{2.37861}\\
$q_{h}\,[(W.u.)^{\frac{1}{2}}]$  &\multicolumn{2}{c|}{4.54818}&\multicolumn{2}{c|}{4.32131}&\multicolumn{2}{c|}{4.71811}&\multicolumn{2}{c|}{3.81321}&\multicolumn{2}{c|}{4.34301}&\multicolumn{2}{c|}{4.57177}\\
$q_{anh}\,[(W.u.)^{\frac{1}{2}}]$&\multicolumn{2}{c|}{1.65546}&\multicolumn{2}{c|}{0.97468}&\multicolumn{2}{c|}{2.64170}&\multicolumn{2}{c|}{1.48316}&\multicolumn{2}{c|}{0}      &\multicolumn{2}{c|}{0}\\
\hline
\end{tabular}}
\end{center}
\end{table}
\clearpage
\renewcommand{\theequation}{6.\arabic{equation}}
\section{Conclusions}
\label{sec:level6}
The present paper consider the CSM approach in two extremes of small and large deformations. Thus, the matrix elements of the model Hamiltonian as well as of the E2 transition operator between the angular momentum projected states modeling the members of the ground, beta and gamma bands, are alternatively expanded in power series of $x (=d^2)$ and $1/x$. As a result the excitation energies in the three bands are expressed analytically as ratios of polynomials in $x$ and $1/x$ respectively with the coefficients depending on angular momentum. Concerning the matrix elements of the E2 transition operator, for small deformation they are, with a few exceptions, linear functions in $d$, the expansion coefficients being rational functions of the angular momentum. 
In the large deformation regime  the whole angular dependence of the mentioned matrix elements is contained by a Clebsch Gordan coefficient which is accompanied by a factor depending on $d$ for intraband and independent of deformation for interband transitions.

This simple description is used to describe the available data for 42 nuclei exhibiting various symmetries like SU(5), O(6), SU(3), triaxial shapes.
The results are in a good agreement with the corresponding experimental data for both excitation energies for the three bands and the transition probabilities. Note that for all symmetries mentioned above we use a sole Hamiltonian and a sole set of projected states. The distinct features of each symmetry are obtained by a specific deformation parameter and  structure coefficients. Changing the nucleus under consideration the coefficients are not changing chaotically but obey a certain rule expressed by their dependence on $A+(N-Z)/2$. In fact this is a measure of the predictability power of the CSM approach. As shown for the Gd isotopes CSM describes not only the nuclei corresponding to a certain symmetry but also those corresponding to the transition between them including the critical nucleus.

Comparing CSM with the Liquid Drop Model  (LDM), one may say that CSM is a highly anharmonic model while LDM has a harmonic structure. However as we mentioned before in the large deformation situation the CSM wave functions are similar to those characterizing  LDM in the strong coupling limit. Another successful anharmonic model was proposed by Gneus and Greiner but that uses a large number of parameters and moreover the quadrupole conjugate momenta contribute to the Hamiltonian only through the quadratic terms. Moreover energies are obtained by diagonalization procedure in a spherical basis which may encounter convergence difficulties for large deformations. By contrast CSM projects, over angular momentum, states from a coherent state and two  orthogonal polynomial excitations and consequently is especially realistic for the well deformed nuclei. This feature is actually confirmed by the application from this paper were the transuranic nuclei spectra are obtained with a high accuracy.

CSM accounts for features which are complementary to those described by IBA. Indeed CSM's  model Hamiltonian is not a boson number conserving Hamiltonian.
Moreover while IBA uses a space of states with limited number of bosons, CSM states covers the whole boson space since they are projected from infinite series of bosons. Due to this feature the IBA approach is concerned with the description of low lying states with angular momentum not exceeding $12^+$
and with a moderate deformation. By contrast, CSM works quite well for high spin states  (in Fig.3 energies for states with $J\le 32$ are shown).
CSM was applied for the description of the triaxial nuclei \cite{UveRaFa} and the results were compared with those obtained within the Vibration Rotation Model \cite{GrFa}. Recently a more extensive study of triaxial nuclei with CSM has been performed \cite{RaBu011} and the results were compared with those produced by a solvable model.

The results of the present paper are quite encouraging for continuing the study of CSM to unveil new virtues suitable to describe even more complex experimental data.
  
{\bf Acknowledgment.} This work was supported by the Romanian Ministry for Education Research Youth and Sport through the CNCSIS project ID-1038/2008.

\renewcommand{\theequation}{A.\arabic{equation}}
\section{Appendix A}
\label{sec:levelA}
\setcounter{equation}{0}
Coefficients of the near vibrational energy formulas are given as follows:
\begin{eqnarray}
\sum_{k=0}^{3}Q_{J,k}^{(\gamma,0)}x^{k}&=&\frac{(J+1)(J+2)(2J+3)}{6(J-1)}+\frac{(J+2)(7J^{2}+7J-24)}{(J-1)(2J+3)}x\nonumber\\
&+&\frac{3}{2}\frac{20J^{4}+85J^{3}+85J^{2}+38J+42}{(J-1)(2J+3)^{2}(2J+5)}x^{2}\nonumber\\
&+&\frac{9(J+1)(14J^{4}+84J^{3}+108J^{2}-122J-204)}{(J-1)(2J+3)^{3}(2J+5)(2J+7)}x^{3},
\end{eqnarray}
\begin{eqnarray}
\sum_{k=0}^{3}Q_{J,k}^{(\gamma,1)}x^{k}&=&(J+2)^{2}+\frac{9J^{3}+22J^{2}-10J-6}{(2J+1)(2J+3)}x\nonumber\\
&+&\frac{3J(12J^{3}+53J^{2}+81J+22)}{(2J+1)^{2}(2J+3)(2J+5)}x^{2}\nonumber\\
&+&\frac{9J(3J+5)}{(2J+1)^{2}(2J+3)(2J+5)}x^{3},
\end{eqnarray}
\begin{eqnarray}
\sum_{k=0}^{3}R_{J,k}^{(\gamma,0)}x^{k}&=&\frac{1}{12(J-1)}(J-2)(J+1)(J+2)(2J+3)+\nonumber\\
&+&\frac{44J^{5}+199J^{4}+67J^{3}-748J^{2}-948J-144}{12(J-1)(J+2)(2J+3)}x\nonumber\\
&+&\Bigg[\frac{1}{J+2}(22J^{4}+59J^{3}-J^{2}+82J+288)+\nonumber\\
&+&\frac{3(J+1)}{(2J+3)(2J+5)}(2J^{4}+4J^{3}-5J^{2}-7J-24)\Bigg]\nonumber\\
&\times &\frac{x^{2}}{4(J-1)(2J+3)}\nonumber\\
&+&\Bigg[\frac{9(J+1)(34J^{4}+196J^{3}+99J^{2}-959J-1200)}{(2J+3)^{2}(2J+5)(2J+7)}+\nonumber\\
&+&\frac{39J^{3}+81J^{2}-54J-120}{J+2}\Bigg]\frac{x^{3}}{4(J-1)(2J+3)},
\end{eqnarray}
\begin{eqnarray}
\sum_{k=0}^{3}R_{J,k}^{(\gamma,1)}x^{k}&=&\frac{1}{2}(J-1)(J+2)^{2}+\frac{11J^{4}+24J^{3}-43J^{2}-100J-42}{2(2J+1)(2J+3)}x\nonumber\\
&+&\frac{9J^{3}+25J^{2}+38J+3}{(2J+1)(2J+3)}x^{2}+\frac{9}{2}\frac{J(J-1)^{2}(J+2)}{(2J+1)^{2}(2J+3)(2J+5)}x^{2}\nonumber\\
&+&3\frac{J(12J^{3}+41J^{2}+36J+31)}{(2J+1)^{2}(2J+3)(2J+5)}x^{3},
\end{eqnarray}
\begin{eqnarray}
\sum_{k=1}^{3}U_{J,k}^{(\gamma,0)}x^{k}&=&\frac{4J(J+1)(J+2)}{15(J-1)}x-\frac{8(J+1)(J+2)(J-3)}{5(J-1)(2J+3)}x^{2}\nonumber\\
&+&\frac{12}{5}\frac{4J^{4}+23J^{3}+5J^{2}-110J-120}{(J-1)(2J+3)^{2}(2J+5)}x^{3},
\end{eqnarray}
\begin{equation}
\sum_{k=2}^{3}U_{J,k}^{(\gamma,1)}x^{k}=\frac{16}{5}x^{2}-\frac{48}{5}\frac{J}{(J+1)(2J+3)}x^{3},
\end{equation}
\begin{eqnarray}
\sum_{k=0}^{3}Q_{J,k}^{(\beta)}x^{k}&=&3J+10+\frac{3}{7}\frac{17J+15}{2J+3}x+27\frac{(J+1)(J+2)}{(2J+3)^{2}(2J+5)}x^{2}\nonumber\\
&+&81\frac{(J+1)(J+2)}{(2J+3)^{3}(2J+5)(2J+7)}x^{3},
\end{eqnarray}
\begin{eqnarray}
\sum_{k=0}^{3}R_{J,k}^{(\beta)}x^{k}&=&\frac{3}{2}J^{2}+14J+30+\frac{72J^{2}+403J+180}{14(2J+3)}x\nonumber\\
&+&\frac{3}{14}\frac{160J^{3}+1333J^{2}+2847J+1680}{(2J+3)^{2}(2J+5)}x^{2}\nonumber\\
&+&\frac{27}{14}\frac{(J+1)(J+2)(48J^{2}+240J+427)}{(2J+3)^{3}(2J+5)(2J+7)}x^{3},
\end{eqnarray}
\begin{eqnarray}
\frac{35}{6}\sum_{k=0}^{3}U_{J,k}^{(\beta)}x^{k}&=&2J(2J+3)+\frac{-12J^{2}+34J+75}{2J+3}x\nonumber\\
&+&\frac{2(52J^{3}+313J^{2}+606J+378)}{2J+3)^{2}(2J+5)}x^{2}\nonumber\\
&-&216\frac{(J+1)(J+2)(J^{2}+5J+5)}{(2J+3)^{3}(2J+5)(2J+7)}x^{3},
\end{eqnarray}
\begin{eqnarray}
\sum_{k=0}^{3}V_{J,k}^{(\beta)}x^{k}&=&\frac{3}{5}(9J^{2}+60J+100)+\frac{18}{35}\frac{51J^{2}+236J+150}{2J+3}x\nonumber\\
&+&\frac{27}{245}\left[123+119\frac{4J^{3}+37J^{2}+78J+42}{(2J+3)^{2}(2J+5)}\right]x^{2}\nonumber\\
&+&\frac{16524}{35}\frac{(J+1)(J+2)(J^{2}+5J+7)}{(2J+3)^{3}(2J+5)(2J+7)}x^{3},
\end{eqnarray}
\begin{eqnarray}
\sum_{k=1}^{3}B_{J,k}^{(\beta)}x^{k}&=&\frac{96}{175}\left[J^{2}x-\frac{2J(J+2)}{2J+3}x^{2}+\frac{20J^{3}+107J^{2}+192J+117}{(2J+3)^{2}(2J+5)}x^{3}\right],\\
\sum_{k=0}^{3}Z_{J,k}^{(\beta)}x^{k}&=&\frac{48}{5\sqrt{70}}\Bigg[-(3J^{2}+46J+120)+\left(Q_{J,0}^{(\beta)}-\frac{72J^{2}+649J+360}{7(2J+3)}\right)x\nonumber\\
&+&\left(Q_{J,1}^{(\beta)}-\frac{3}{7}\frac{160J^{3}+1711J^{2}+3981J+2436}{(2J+3)^{2}(2J+5)}\right)x^{2}\nonumber\\
&+&\left(Q_{J,2}^{(\beta)}-\frac{27}{7}\frac{(J+1)(J+2)(48J^{2}+240J+553)}{(2J+3)^{3}(2J+5)(2J+7)}\right)x^{3}\Bigg],
\end{eqnarray}
\begin{eqnarray}
\sum_{k=1}^{3}T_{J,k}x^{k}&=&242\frac{J(J+1)(J+2)}{J-1}\left[\frac{x}{6}-\frac{x^{2}}{2J+3}+\frac{3}{2}\frac{4J+7}{(2J+3)^{2}(2J+5)}x^{3}\right],\\
\sum_{k=0}^{3}X_{J,k}x^{k}&=&\frac{(J+1)(J+2)(2J+3)}{6(J-1)}+\frac{3(4J^{3}+11J^{2}-11J-34)}{(J-1)(2J+3)}x\nonumber\\
&+&\Bigg[-\frac{88J^{5}+196J^{4}-460J^{3}-973J^{2}-585J-396}{4(J-1)(2J+3)^{2}(2J+5)}\nonumber\\
&+&\frac{22J^{3}+15J^{2}-31J+144}{4(J-1)(2J+3)}\Bigg]x^{2}\nonumber\\
&+&\frac{9}{4}\Bigg[\frac{(J+1)(-28J^{5}-24J^{4}+639J^{3}+1176J^{2}-1559J-2844)}{(J-1)(2J+3)^{3}(2J+5)(2J+7)}\nonumber\\
&+&\frac{14J^{5}+81J^{4}+137J^{3}-13J^{2}-191J-108}{(J-1)(2J+3)^{3}(2J+5)}\Bigg]x^{3}.
\end{eqnarray}

The term $\Delta E_{J}$ accounts for the interaction between the states $\phi_{JM}^{g}$ and $\phi_{JM}^{\gamma}$. In the near vibrational regime this has the expression
\begin{equation}
\Delta E_{J}=A_{1}\frac{\sum_{k=0}^{3}T_{J,k}x^{k}}{\sum_{k=0}^{3}\left(22X_{J,k}+5U_{J,k}^{(\gamma,0)}\right)x^{k}},
\end{equation}
with coefficients $T_{J,k}$ and $X_{J,k}$  given above.

\renewcommand{\theequation}{B.\arabic{equation}}
\section*{Appendix B}
\label{sec:levelB}
\setcounter{equation}{0}

\begin{align}
P_{J}^{\beta}=&\frac{12}{5}+\frac{171}{35}x-\frac{6}{5x}+\left(\frac{3}{5x}+\frac{1}{x^{2}}+\frac{13}{5x^{3}}\right)J(J+1)-\frac{1}{15x^{3}}J^{2}(J+1)^{2},\\
S_{J}^{\beta}=&\frac{2}{35}\left[1917x^{2}+5946x+759-\frac{1937}{x}\right]\nonumber\\
&+\frac{J(J+1)}{35}\left[1125+\frac{2537}{x}+\frac{14365}{3x^{2}}+\frac{25181}{3x^{3}}\right]\nonumber\\
&-\frac{J^{2}(J+1)^{2}}{7x^{2}}\left(8+\frac{1937}{45x}\right),\\
F_{J}^{\beta}=&\frac{54}{1225}\bigg[-\frac{406}{3}+\frac{1083}{2}x^{2}+826x-\frac{833}{3x}+\bigg(133+\frac{714}{3x}+\frac{4403}{9x^{2}}\nonumber\\
&+\frac{10829}{18x^{3}}\bigg)J(J+1)-\frac{119}{18x^{2}}\left(1+\frac{7}{3x}\right)J^{2}(J+1)^{2}\bigg],
\end{align}

For $J=$odd we have
\begin{align}
S_{J}^{\gamma}=&198(J+1)x^{3}+(J+1)(-66J+368)x^{2}+(J+1)\bigg(77J^{2}-\frac{335}{3}J-\frac{188}{3}\bigg)x\nonumber\\
&+\frac{1}{9}(-165J^{4}+1023J^{3}+635J^{2}-2219J-2046)\nonumber\\
&+\frac{(J-1)}{27x}(33J^{4}-688J^{3}+4549J^{2}+7098J-396)\nonumber\\
&+\frac{11J(J-1)}{27x^{2}}(2J^{4}+3J^{3}-162J^{2}+499J+1296)\nonumber\\
&-\frac{11}{27x^{3}}J(J-1)^{3}(J-2)(J^{2}-J-39),\\
P_{J}^{\gamma}=&9(J+1)x^{2}-(3J-4)(J+1)x+\frac{1}{3}(J+1)(6J^{2}-7J-7)\nonumber\\
&+\frac{1}{9x}(J-1)(-3J^{3}+21J^{2}+28J-6)\nonumber\\
&+\frac{1}{x^{2}}J(J-1)(-\frac{1}{9}J^{3}-\frac{17}{27}J^{2}+\frac{152}{27}J+\frac{20}{3})\nonumber\\
&+\frac{J(J-1)^{3}}{27x^{3}}(J^{2}-J-39).
\end{align}

For $J=$even we have
\begin{gather}
S_{J}^{\gamma}=\sum_{k=0}^{4}U^{(k)}J^{k}(J+1)^{k},\,\,\,\,P_{J}^{\gamma}=\sum_{k=0}^{3}V^{(k)}J^{k}(J+1)^{k},
\end{gather}
where
\begin{align}
U^{(0)}=&-396x^{3}-736x^{2}+\frac{376}{3}x+\frac{1364}{3}-\frac{88}{3x},\nonumber\\
U^{(1)}=&198x^{3}+368x^{2}-\frac{584}{3}x-\frac{3835}{9}-\frac{4655}{9x}-\frac{1056}{x^{2}}+\frac{572}{9x^{3}},\nonumber\\
U^{(2)}=&66x+\frac{1037}{9}+\frac{5702}{27x}+\frac{18847}{54x^{2}}+\frac{451}{6x^{3}},\nonumber\\
U^{(3)}=&-\frac{47}{54x}-\frac{49}{9x^{2}}+\frac{539}{162x^{3}},\nonumber\\
U^{(4)}=&-\frac{11}{81x^{3}},\\
V^{(0)}=&-18x^{2}-8x+\frac{14}{3}-\frac{4}{3x},\nonumber\\
V^{(1)}=&9x^{2}+4x-\frac{16}{3}-\frac{53}{9x}-\frac{40}{3x^{2}}+\frac{26}{9x^{3}},\nonumber\\
V^{(2)}=&\frac{3}{2}+\frac{22}{9x}+\frac{169}{27x^{2}}+\frac{113}{54x^{3}},\nonumber\\
V^{(3)}=&-\frac{7}{54x^{2}}-\frac{1}{18x^{3}}.
\end{align}

The factors $T_{J}^{n,\beta}$, with n=4,5, involved in the equation determining the excitation energies in the beta band have the following expression:
\begin{eqnarray}
T_J^{4,\beta}&=&
\frac{171}{35}x^{2}+\frac{195}{7}x+\frac{321}{35}-\frac{361}{35x}-\frac{1949}{105x^{2}}-\frac{1591}{45x^{3}}\nonumber\\
&&+J(J+1)\Bigg(\frac{99}{70}+\frac{361}{70x}+\frac{1973}{210x^{2}}+\frac{559}{30x^{3}}\Bigg)\nonumber\\
&&-\frac{1}{54x^{3}}\Bigg(\frac{129}{5}+\frac{108}{35}x\Bigg)J^{2}(J+1)^{2},\nonumber\\
T_J^{5,\beta}&=&\Bigg[9x^{3}+24x+16+\frac{104}{3x}+\frac{74}{3x^{2}}+\left(6x-8-\frac{18}{x}-\frac{13}{x^{2}}\right)J(J+1)\nonumber\\
&&+\frac{x+1}{3x^{2}}J^{2}(J+1)^{2}\Bigg].
\end{eqnarray}
\renewcommand{\theequation}{C.\arabic{equation}}
\section*{Appendix C}
\label{sec:levelC}
\setcounter{equation}{0}
The exact m.e. of the harmonic part of the quadrupole transition operator \cite{Raduta1} can be expressed in terms of  the projected state norms:
\begin{eqnarray}
\langle\phi_{J}^{g}||Q_{2}^{h}||\phi_{J'}^{g}\rangle&=&q_{h}dC^{J'\,2\,J}_{0\,0\,0}\left[\frac{2J'+1}{2J+1}\frac{N_{J'}^{g}}{N_{J}^{g}}+\frac{N_{J}^{g}}{N_{J'}^{g}}\right],\\
\langle\phi_{J}^{\beta}||Q_{2}^{h}||\phi_{J'}^{\beta}\rangle&=&q_{h}dC^{J'\,2\,J}_{0\,0\,0}\left[\frac{N_{J}^{\beta}}{N_{J'}^{\beta}}+\frac{18}{5}\frac{N_{J}^{\beta}N_{J'}^{\beta}}{(N_{J'}^{g})^{2}}+\frac{2J'+1}{2J+1}\left(\frac{N_{J'}^{\beta}}{N_{J}^{\beta}}+\frac{18}{5}\frac{N_{J}^{\beta}N_{J'}^{\beta}}{(N_{J}^{g})^{2}}\right)\right],\\
\langle\phi_{J}^{g}||Q_{2}^{h}||\phi_{J'}^{\beta}\rangle&=&0,\\
\langle\phi_{J}^{\gamma}||Q_{2}^{h}||\phi_{J'}^{g}\rangle&=&q_{h}d N_{J}^{\gamma}\Bigg[\sqrt{\frac{2}{7}}C^{J'\,2\,J}_{0\,2\,2}\frac{1}{N_{J'}^{g}}+2\sum_{J_{1}}\hat{2}\hat{J}C^{2\,J\,J_{1}}_{-2\,2\,0}C^{J'\,2\,J_{1}}_{0\,0\,0}\nonumber\\
&&\times W(22JJ';2J_{1})\frac{2J'+1}{2J_{1}+1}\frac{N_{J'}^{g}}{\left(N_{J_{1}}^{g}\right)^{2}}\Bigg],\\
\langle\phi_{J}^{\beta}||Q_{2}^{h}||\phi_{J'}^{\gamma}\rangle
&=&q_{h}N_{J}^{\beta}N_{J'}^{\gamma}(2J'+1)\frac{6}{7\sqrt{5}}\Bigg\{C^{J'\,2\,J}_{2-2\,0}\frac{1}{2J+1}\left[3\left(\frac{2}{7}d^{2}-1\right)\left(N_{J}^{g}\right)^{-2}+\frac{5}{3}\left(N_{J}^{\beta}\right)^{2}\right]\nonumber\\
&&-2d^{2}C^{J'\,2\,J}_{2\,0\,2}\sum_{J_{1}}\frac{1}{2J_{1}+1}C^{J\,2\,J_{1}}_{0\,0\,0}C^{J\,2\,J_{1}}_{2-2\,0}\left(N_{J_1}^{g}\right)^{2}\Bigg\},\\
\langle\phi_{J}^{\gamma}||Q_{2}^{h}||\phi_{J'}^{\gamma}\rangle&=&q_{h}\left[1+\frac{\hat{J'}}{\hat{J}}(-)^{J'-J}(J'\leftrightarrow J)\right]\langle\phi_{J}^{\gamma}||b||\phi_{J'}^{\gamma}\rangle,\\
\langle\phi_{J}^{\gamma}||b||\phi_{J'}^{\gamma}\rangle&=&d(2J'+1)N_{J}^{\gamma}N_{J'}^{\gamma}\Bigg\{\frac{1}{2J+1}C^{J'\,2\,J}_{2\,0\,2}\left(N_{J}^{\gamma}\right)^{-2}+\sum_{J_{1}}C^{J'\,2\,J_{1}}_{2-2\,0}W(J'2J_{1}2;J2)\nonumber\\
&&\times\left[2\sqrt{\frac{2}{7}}\frac{\hat{2}}{\hat{J_{1}}}C^{J_{1}\,2\,J}_{0-2-2}\left(N_{J_{1}}^{\gamma}\right)^{-2}+20\sum_{J_{2}}\frac{\hat{J_{1}}}{\hat{J_{2}}}C^{J_{1}\,2\,J_{2}}_{0\,0\,0}C^{J_{2}\,2\,J}_{0-2-2}W(J2J_{2}2;J_{1}2)\left(N_{J_{2}}^{\gamma}\right)^{-2}\right]\Bigg\}.\nonumber\\
\end{eqnarray}

The exact expressions for  the m.e. of the anharmonic quadrupole transition operator are following \cite{Raduta1}:
\begin{eqnarray}
\langle\phi_{J}^{g}||Q_{2}^{anh}||\phi_{J'}^{g}\rangle&=&-q_{1}d^{2}C^{J'\,2\,J}_{0\,0\,0}\left[\frac{2J'+1}{2J+1}\frac{N_{J'}^{g}}{N_{J}^{g}}+\frac{N_{J}^{g}}{N_{J'}^{g}}\right],\\
\langle\phi_{J}^{\beta}||Q_{2}^{anh}||\phi_{J'}^{\beta}\rangle&=&-q_{1}d^{2}C^{J'\,2\,J}_{0\,0\,0}\left[\frac{N_{J}^{\beta}}{N_{J'}^{\beta}}+\frac{2J'+1}{2J+1}\frac{N_{J'}^{\beta}}{N_{J}^{\beta}}\right],\\
\langle\phi_{J}^{g}||Q_{2}^{anh}||\phi_{J'}^{\beta}\rangle&=&-6\sqrt{\frac{1}{5}}q_{1}dC^{J'\,2\,J}_{0\,0\,0}\frac{N_{J}^{g}N_{J'}^{\beta}}{N_{J'}^{g}},\\
\langle\phi_{J}^{\gamma}||Q_{2}^{anh}||\phi_{J'}^{g}\rangle&=&q_{1}N_{J}^{\gamma}N_{J'}^{g}\Bigg[2\left(N_{J'}^{g}\right)^{-2}C^{J'\,2\,J}_{0\,2\,2}\left(1+\frac{2}{7}d^{2}\right)\nonumber\\
&&+20d^{2}\hat{J'}\sum_{J_{1}J_{2}}\hat{J_{2}}C^{J_{1}\,J_{2}\,J}_{0-2-2}C^{J_{1}\,2\,J'}_{0\,0\,0}C^{2\,2\,J_{2}}_{0\,2\,2}T_{J_{1}J_{2}}^{JJ'}\Bigg],\\
\langle\phi_{J}^{\beta}||Q_{2}^{anh}||\phi_{J'}^{\gamma}\rangle&=&\frac{q_{anh}}{q_h}\left[\frac{6}{\sqrt{5}}\langle\phi_{J}^{g}||Q_{h}||\phi_{J'}^{\gamma}\rangle\frac{N_{J}^{\beta}}{N_{J}^{g}}\right.\nonumber\\
&&\left.+2d\hat{2}\sum_{J_{1}}\hat{J}_{1}C^{J_{1}\,0\,J}_{0\,0\,0}W(22JJ';2J_{1})\langle\phi_{J_{1}}^{\beta}||Q_{h}||\phi_{J'}^{\gamma}\rangle\frac{N_{J}^{\beta}}{N_{J_{1}}^{\beta}}\right].\\
\langle\phi_{J}^{\gamma}||Q_{2}^{anh}||\phi_{J'}^{\gamma}\rangle&=&q_{anh}\left[1+\frac{\hat{J'}}{\hat{J}}(-)^{J'-J}(J'\leftrightarrow J)\right]\langle\phi_{J}^{\gamma}||(bb)_{2}||\phi_{J'}^{\gamma}\rangle,\\
\langle\phi_{J}^{\gamma}||(bb)_{2}||\phi_{J'}^{\gamma}\rangle&=&N_{J}^{\gamma}N_{J'}^{\gamma}\Bigg\{-\sqrt{\frac{2}{7}}C^{J'\,2\,J}_{2\,0\,2}d^{2}(N_{J'}^{\gamma})^{-2}\nonumber\\
&&+20d^{2}\sqrt{\frac{2}{7}}\hat{J'}\sum_{J_{1}J_{2}}\hat{J_{2}}C^{J_{1}\,2\,J'}_{0-2-2}C^{J_{1}\,J_{2}\,J}_{0-2-2}C^{2\,2\,J_{2}}_{0\,2\,2}T^{JJ'}_{J_{1}J_{2}}\nonumber\\
&&+40d^{2}\hat{2}\hat{J'}\hat{J}\sum_{J_{1}J_{2}J_{3}}\hat{J_{2}}C^{2\,2\,J_{3}}_{2\,0\,2}C^{J\,J_{2}\,J_{3}}_{2\,0\,2}C^{J_{1}\,2\,J_{2}}_{0\,0\,0}C^{J_{1}\,2\,J'}_{0-2-2}S^{JJ'}_{J_{1}J_{2}J_{3}}\Bigg\}.
\end{eqnarray}

In the above equations, the following notations were used:
\begin{eqnarray}
T_{J_{1}J_{2}}^{JJ'}&=&W(2222;2J_{2})W(J'2J_{1}J_{2};J2)\left(N_{J_{1}}^{g}\right)^{-2},\\
S^{JJ'}_{J_{1}J_{2}J_{3}}&=&W(2222;2J_{3})W(J_{3}2JJ';2J_{2})W(22J'J_{2};2J_{1})\left(N_{J_{1}}^{g}\right)^{-2}.
\end{eqnarray}

\end{document}